\documentclass[useAMS,usenatbib]{mnras}

\usepackage{graphicx}
\usepackage{color}
\usepackage{times}
\usepackage{natbib}
\usepackage{setspace}

\usepackage{amsmath}
\usepackage{amsbsy}

\newif\ifAMStwofonts
\AMStwofontstrue
\definecolor{red}{rgb}{1,0.,0.}

\newcommand{\K}{{\rm K}}

\newcommand{\hi}{H{\sc i}}

\newcommand{\arepo}{{\sc arepo}}

\newcommand{\pc}{{\rm pc}}
\newcommand{\kpc}{{\rm kpc}}
\newcommand{\Mpc}{{\rm Mpc}}

\newcommand{\kms}{{\rm km~s^{-1}}}
\newcommand{\cm}{{\rm cm}}

\newcommand{\gsim}{\,\lower.7ex\hbox{$\;\stackrel{\textstyle>}{\sim}\;$}}
\newcommand{\lsim}{\,\lower.7ex\hbox{$\;\stackrel{\textstyle<}{\sim}\;$}}
\newcommand{\eagle}{{\sc eagle}}

\newcommand{\FM}[1]{#1}

\defcitealias{Planck2014}{Planck Collaboration XVI} 
\defcitealias{Bahe2016}{B16}
\defcitealias{Leroy2008}{L08}
\defcitealias{Gnedin2011}{GK11}

\newcommand{\planckpap}{\citetalias{Planck2014} \citeyearpar{Planck2014}}

\ifpdf
  \DeclareGraphicsExtensions{.pdf, .jpg}
\else
  \DeclareGraphicsExtensions{.eps}
\fi

\title[\hi\ discs in the Auriga simulations] 
{Properties of \hi\ discs in the Auriga cosmological simulations}

\author[F. Marinacci et al.]
{Federico Marinacci$^1$\thanks{E-mail: fmarinac@mit.edu}, Robert J.~J. Grand$^{2,3}$, 
  R\"udiger Pakmor$^2$, Volker Springel$^{2,3}$,\newauthor Facundo A. G\'omez$^4$,
  Carlos S. Frenk$^5$ and Simon D.~M. White$^4$
  \vspace*{0.2cm}  \\
  $^1$Kavli Institute for Astrophysics and Space Research, 
  Massachusetts Institute of Technology, Cambridge, MA 02139, USA\\
  $^2$Heidelberger Institut f\"ur Theoretische Studien,
Schloss-Wolfsbrunnenweg 35, 69118 Heidelberg, Germany \\
  $^3$Zentrum f\"ur Astronomie der Universit\"at Heidelberg, ARI, M\"onchhofstrasse
12-14, 69120 Heidelberg, Germany\\
  $^4$Max-Planck-Institut f\"ur Astrophysik, Karl-Schwarzschild-Str. 1, D-85748, 
Garching, Germany\\
  $^5$ Institute for Computational Cosmology, Department of Physics, Durham University, 
  South Road, Durham, DH1 3LE, UK
}

\date{Accepted 2016 December 21. Received 2016 December 21; in original form 2016 October 05}

\begin{document}

\pagerange{\pageref{firstpage}--\pageref{lastpage}}
\pubyear{2016}

\maketitle

\label{firstpage}

\begin{abstract} 
  We analyse the properties of the \hi\ gas distribution in the Auriga
  project, a set of magnetohydrodynamic cosmological simulations
  performed with the moving-mesh code \arepo\ and a physics model for
  galaxy formation that succeeds in forming realistic late-type
  galaxies in the 30 Milky Way-sized haloes simulated in this
  project. We use a simple approach to estimate the neutral hydrogen
  fraction in our simulation set, which treats low-density and
  star-forming gas separately, and we explore two different
  prescriptions to subtract the contribution of molecular hydrogen
  from the total \hi\ content. The \hi\ gas in the vast majority of
  the systems forms extended discs although more disturbed
  morphologies are present. Notwithstanding the general good agreement with
  observed \hi\ properties -- such as radial profiles and the 
  mass-diameter relation -- the Auriga galaxies are systematically larger 
  and more gas-rich than typical nearby galaxies. Interestingly, the amount 
  of \hi\ gas outside the disc plane correlates with the star formation rate, 
  consistent with a picture where most of this extra-planar \hi\ gas 
  originates from a fountain-like flow. Our findings are robust with 
  respect to the different assumptions adopted for computing the 
  molecular hydrogen fraction and do not vary significantly 
  over a wide range of numerical resolution. The \hi\ modelling 
  introduced in this paper can be used in future work to build artificial 
  interferometric \hi\ data cubes, allowing an even closer comparison of 
  the gas dynamics in simulated galaxies with observations.
\end{abstract}


\section{Introduction} \label{sec:intro}

Cold gas plays a key role in the evolution of galaxies, in
particular in those that are still actively forming stars like our own
Milky Way. Indeed, the presence of a cold gas reservoir is necessary for a
galaxy to keep forming stars \citep[e.g.][and references
therein]{Sancisi2008}.  Moreover, several pieces of evidence
\citep{Twarog1980, Chiappini1997, Chiappini2001, Aumer2009} indicate
that some process of accretion, replacing the gas consumed by star
formation with fresh (and possibly metal-poor) material, must be
ongoing for a substantial fraction of the lifetime of a star-forming
galaxy.

A very important component of this cold gas phase is represented by
atomic hydrogen (\hi). Radio observations of \hi\ in nearby galaxies
now date back several decades \citep{Roberts1975} and have 
been instrumental to study the kinematics
\citep[e.g.][]{Verheijen2001}, structure \citep[e.g.][]{Broeils1997}
and gaseous content \citep[e.g.][]{Catinella2013} of (late-type)
galaxies \citep[see also][for an updated review on \hi\
surveys]{Giovanelli2016}. As the sensitivity of these observations
improved with time \citep{Heald2011, Kalberla2005, Wang2013}, it has
become possible to detect fainter and fainter \hi\ structures, which
has not only allowed a better census of the total \hi\ content in
spiral galaxies but also elucidated the interactions they have with
their environment and their importance for galactic evolution. For
example, the detection of extra-planar gas layers (i.e.  low-column
density \hi\ gas that extends well outside the plane of the galactic
disc, see e.g. \citealt{Oosterloo2007}) and the study of their
kinematics \citep[][and references therein]{Fraternali2009} has
revealed the interplay between the disc and the circumgalactic
medium in late-type galaxies \citep{Fraternali2008, Marinacci2010,
  Marinacci2011}. This interaction is mediated by a so-called
galactic fountain \citep{Shapiro76, HouckB90} -- which in the Milky
Way manifests itself through intermediate-velocity (IVC) and high-velocity
(HVC) clouds \citep{Fraternali2015, Marasco2011, Wakker1997,
  Wakker2004} -- and provided indirect evidence of how star formation
can be sustained for a Hubble time in these systems
\citep{Fraternali2014}.

The reproduction of the content and distribution of \hi\ gas in
theoretical models of galaxy formation is of primary importance, given
its key role in shaping galactic evolution. In fact, the
numerous constraints posed by such observations represent a
challenging test for any galaxy formation model. This is particularly
true in hydrodynamical simulations, in which, due to the huge dynamic
range in spatial scales that needs to be covered, many of the physical
processes crucial for forming realistic objects are included in an
approximate sub-grid fashion. The latter may in principle introduce a
substantial uncertainty in the final results, and also an explicit
dependence on the details of the numerical implementation of the
sub-grid processes, calling for a validation against observations.

Some work in this direction has already been carried out, both in
idealized set-ups and in full cosmological settings. These efforts
mostly focused on the dependence on supernova feedback -- a key
process in galaxy formation models, especially for galaxies with
virial masses up to $\sim 10^{12}\,{\rm M}_{\odot}$ -- of the extra
planar \hi\ distribution and kinematics in idealized Milky Way type
galaxies \citep{Marasco2015}, and of the global \hi\ properties
in full cosmological set-ups \citep[e.g.][]{Dave2013, Popping2009,
  Rahmati2013, Walker2014}.  However, the majority of this earlier
cosmological work featured a simulated galaxy population with
properties that are in substantial tension with observations, even when
matching the large-scale \hi\ observational constraints.  In
particular, the simulated galaxies invariably exhibited too high
stellar masses and lacked a well-defined disc component, due to an
overly efficient cooling of baryons at the centre of dark matter
haloes (the so-called `over-cooling' problem, e.g.~\citealt{Navarro1997}).

It was not until recently that this issue has been successfully
addressed, essentially by invoking a more efficient coupling of the
feedback energy -- both stellar and, at the high
$\gsim 10^{12}\, {\rm M}_{\odot}$ mass end, from active galactic
nuclei -- with the cooling gas. This more effective coupling can be
very difficult to numerically resolve in full-scale cosmological
simulations, and different groups have explored various approaches to
address this problem. Notwithstanding these technical difficulties,
there are now several simulation methodologies that are able to
produce realistic looking galaxies in hydrodynamic cosmological
simulations both in uniformly-sampled boxes -- as for instance in the
Illustris \citep{Vogelsberger2014b, Vogelsberger2014a} and \eagle\
\citep{Schaye2015} simulation projects -- and in configurations that focus on 
the formation of a single object, \FM{commonly a Milky Way-sized galaxy 
\citep[e.g.][]{Aumer2013b, Colin2016, Guedes2011, Marinacci2014a, Stinson2013,
Wang2015}, or a region of prespecified density \citep{McCarthy2012}}.

The availability of a new generation of cosmological simulations
yielding realistic galaxies implies that it is now possible to study
their \hi\ content and kinematics more reliably than possible in the past
(for very recent analysis see \citealt{Crain2016, Maccio2016, Marasco2016}), 
and also to check whether the
increased feedback efficiency, which nominally should significantly
affect the gas properties within and in the vicinity of the star
forming disc, has an effect on the (global) distribution of the \hi\
gas. This has been explored in the context of the \eagle\
collaboration by \citet[][hereafter \citetalias{Bahe2016}]{Bahe2016},
who have analysed several properties of the \hi\ gas -- including
morphologies, radial distribution, gas fractions, \hi\ masses and
sizes -- of the simulated galaxy set. In their study,
\citetalias{Bahe2016} conclude that the increased efficiency of
the stellar feedback and its local character have indeed a large impact on
the morphology of the \hi\ discs, and strongly influences its radial
distribution. Other properties, such as \hi\ masses and sizes, appear
to be less sensitive to the increased (stellar) feedback efficiency.

Similarly to the work of \citetalias{Bahe2016}, in this paper we
present an analysis of the \hi\ properties of a set of 30 cosmological
zoom-in magneto-hydrodynamic (MHD) simulations of disc galaxy formation, named the Auriga
simulations, performed with the moving-mesh code \arepo\ and a
state-of-the-art model of galaxy formation physics largely based on
the one used for the Illustris project
\citep[see][]{Vogelsberger2013}, slightly adapted for zoom-in
simulations of Milky Way type objects \citep{Marinacci2014a}.
Compared to the study of \citetalias{Bahe2016} the zoom-in technique
makes it possible to reach a finer (mass and spatial) resolution
\FM{in Auriga -- by factors of about 36 in mass and 3 in gravitational softening 
length compared to the fiducial resolution of \eagle\ --} and
the relatively large number of simulated haloes allows us to draw
statistically meaningful conclusions about the properties of the
resulting \hi\ discs, albeit limited to halo masses around
$\simeq 10^{12}\, {\rm M}_{\odot}$ which is the range of masses chosen
as Milky Way-like candidate haloes in the Auriga project.  We note
that the \citetalias{Bahe2016} and our simulation set have very 
different treatments for the star-forming gas and stellar feedback,
enabling an interesting comparison on how the numerical implementation
of these aspects impacts the simulated \hi\ distribution.

This paper is structured as follows. In Section~\ref{sec:Auriga}, we
briefly describe the main properties of the simulations that we have
analysed in this study and the numerical techniques employed to carry
them out. Section~\ref{sec:models} presents the method that we have
adopted to estimate the \hi\ content of the Auriga discs, and in
particular gives details on the two prescriptions that we have used to
treat the molecular gas fraction. In Section~\ref{sec:results}, we
illustrate our main results focusing on the structural properties,
scaling relations and gas content of the simulated \hi\ discs. These
findings are compared to observations and with previous simulation
results.  Finally, in Section~\ref{sec:conclusions}, we present our
conclusions.

\section{The Auriga simulations}\label{sec:Auriga}

In this work, we investigate the \hi\ content of the cosmological
hydrodynamical simulation suite Auriga \citep{Grand2016b}. 
This simulation suite is aimed at modelling the formation and
evolution of Milky Way-like galaxies in their proper cosmological
context. Full technical details about the simulation suite can be
found in \citet{Grand2016b}\footnote{For more information about the simulation suite see 
also the project's website \url{http://auriga.h-its.org}}. In the following, we restrict
ourselves to a concise description of the numerical methods adopted to
run the simulations and their main properties.

The Auriga simulations follow a $\Lambda$ cold dark matter cosmology with parameters
$\Omega_{\rm m}$ = $\Omega_{\rm dm}$ + $\Omega_{\rm b}$ = 0.307,
$\Omega_{\rm b}$ = 0.048, $\Omega_{\Lambda}$ = 0.693, and Hubble
constant $H_0 = 100\,h\,\kms \Mpc^{-1}$ = 67.77 $\kms \Mpc^{-1}$,
consistent with the \planckpap\ data release. The candidate haloes
selected for resimulation were identified in a low-resolution
simulation of a periodic cosmological box of side $100\,\Mpc$.
Candidate haloes were selected in a virial mass range around
$10^{12}\, {\rm M}_{\odot}$, consistent with recent determinations of
the MW mass \citep{Wilkinson1999, Sakamoto2003, Battaglia2005,
  Dehnen2006, Li2008, Xue2008, Penarrubia2016}, a mild isolation
criterion -- no object of more than half the mass of the candidate is
allowed to be closer than $1.37\,\Mpc$ -- was enforced at $z = 0$ to
select relatively isolated objects.

\begin{figure*}
\centering
\includegraphics[width=0.24\textwidth]{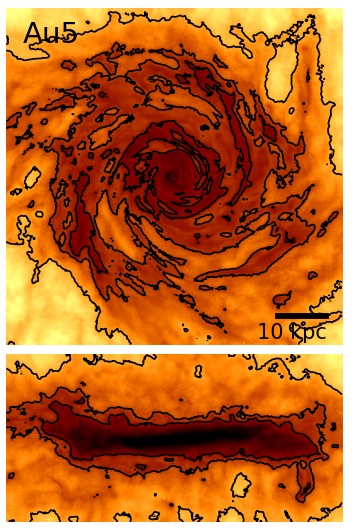}
\includegraphics[width=0.24\textwidth]{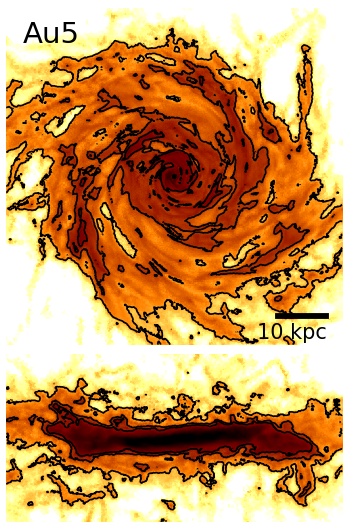}
\includegraphics[width=0.24\textwidth]{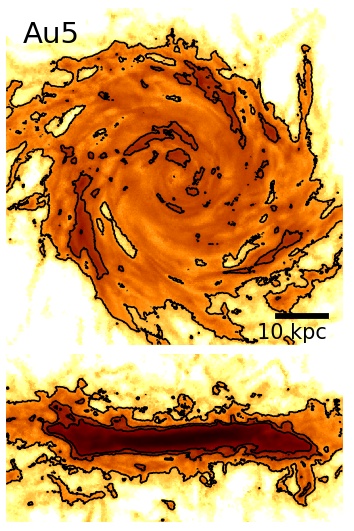}
\includegraphics[width=0.24\textwidth]{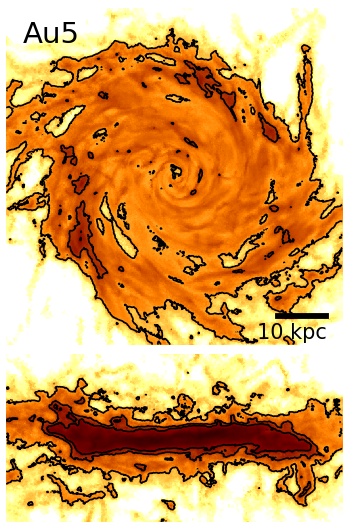}
\caption{\hi\ projections. From left to right: total H density,
  neutral H density (\hi\ + H$_2$), \hi\ density from the
  \citetalias{Leroy2008} pressure prescription, \hi\ density from the
  \citetalias{Gnedin2011} prescription. The size of the projected
  region is 50 kpc on a side for face on projections and $50\times 25$
  kpc for edge-on projections. To align the projections, the spin
  direction of the cold, star-forming gas within 10\% of the virial
  radius of the halo has been used. The colour scale logarithmically
  maps the column densities in the interval
  $[5\times 10^{18}, 10^{22}]\, \cm^{-2}$, and contour levels are
  placed at 10, 5 and 1 ${\rm M}_{\odot}\,\pc^{-2}$ from the inside to the outside.  }
\label{fig:HIprojections}
\end{figure*}

The selected haloes (denoted with the acronym Au $i$, where $i$ is the
halo ID) were simulated multiple times at different resolution levels
by applying the so-called ``zoom-in'' technique, in which the mass
distribution in the Lagrangian region that forms the main halo is
sampled by a large number of resolution elements, whereas
progressively coarser resolution is applied with increasing distance
from the target object. At the fiducial resolution level that we are
going to analyse in the present work (which from now on will be referred
to as level 4), the typical baryonic mass resolution is
$\sim 4 \times 10^4 {\rm M}_{\odot}$ and that of the dark matter
component is $\sim 3 \times 10^5 {\rm M}_{\odot}$. The softening
length of the dark matter particles is kept fixed in comoving space
and allowed to grow in physical space up to a maximum of $369\,\pc$,
while the gas softening length is scaled by the mean radius of the
cell. The maximum physical softening allowed is $1.85 \,\kpc$.

Gas was included in the initial conditions by adopting the same
technique as in \citet[][and references
therein]{Marinacci2014a,Marinacci2014b}. In practice, each particle in
the initial conditions is split into a pair of a dark matter particle and a gas
cell. The phase space coordinates of the pair are chosen such that
its centre of mass and centre of mass velocity are equal to those of
the original dark matter particle. The mass ratio of the pair is set
by the cosmic baryon fraction, and the dark matter particle and gas
cell are separated by a distance equal to half the mean inter-particle
spacing.

The simulation suite was carried out using the moving-mesh,
MHD code \arepo\ \citep{Springel2010}. To solve for
collisionless dynamics, \arepo\ employs a TreePM approach, in which
the gravitational force is split into a long-range and a short-range
component \citep[see][for details]{Springel2005}. The long-range
contribution is calculated on a Cartesian mesh via a Fourier method,
while for the short-range component a gravitational oct-tree
\citep{Barnes1986} is used. The MHD equations
are discretized in a finite-volume approach on a dynamic unstructured
Voronoi tessellation of the simulation domain.  The special trait of
\arepo, distinguishing it from other cosmological finite-volume codes,
is that the mesh-generating points defining the Voronoi tessellation
are allowed to move with the local fluid velocity, thereby giving rise
to a quasi-Lagrangian numerical method which is manifestly
Galilean-invariant. The MHD equations are solved with a second order
Runge-Kutta integration scheme complemented with least square spatial
gradient estimators of primitive variables \citep{Pakmor2016} and the
\citet{Powell1999} eight-wave cleaning scheme to ensure the
$\nabla\cdot\boldsymbol{B} = 0$ constraint.

The Auriga simulations use a galaxy formation physics model that
accounts for the baryonic processes that play a key role in the
formation of late-type galaxies. The model itself is an updated
version of the one used in \citet{Marinacci2014a, Marinacci2014b},
which was specifically developed for large-scale cosmological
simulations and whose free parameters were calibrated against a small
set of observables such as the cosmic star formation history and the
galaxy stellar mass function. For a full description of the
implementation details we refer the reader to
\citet{Vogelsberger2013}. Differences of the present galaxy formation
module with respect to the \citet{Marinacci2014a} implementation are
discussed in \citet{Grand2016b}.

Magnetic fields are included in the initial conditions as in
\citet{Pakmor2014} by seeding a homogeneous $B$-field of $10^{-14}$
(comoving) G along the $z$-direction, which is then amplified due to
the assembly of the main galaxy. The seeding is necessary because, in
ideal MHD, $B$-fields cannot be self-consistently generated when starting
from zero-field initial conditions. We note that previous cosmological
MHD simulations with \arepo\ have shown that the results are
essentially insensitive to the choice of both the intensity and the
direction of the seed field \citep{Pakmor2014}, at least inside
virialized structures \citep{Marinacci2015} where any memory of the
initial conditions is quickly erased due to vigorous, small-scale
dynamo amplification of the field.

\section{Computation of the \hi\ mass} \label{sec:models}

The first step in our analysis is to estimate the amount of neutral atomic 
hydrogen contained in any given cell ($M_{{\rm HI}, 
i}$) of our simulated objects, which can be expressed as follows 
\begin{equation}
M_{{\rm HI}, i} = (1 - f_{{\rm mol},i}) f_{{\rm neutr},i} X_{i} M_{i},
\label{eq:HImass}
\end{equation}
where $M_i$ is the total gas mass contained in a given cell $i$, $X_i$
is the hydrogen mass fraction, $f_{{\rm neutr},i}$ the neutral
hydrogen fraction and $f_{{\rm mol},i}$ the molecular hydrogen
fraction, respectively. $M_i$ and $X_i$ are readily available from the
simulation output, since \arepo\ directly tracks their evolution. The
determination of the neutral H fraction $f_{{\rm neutr},i}$ and of the
molecular H fraction $f_{{\rm mol},i}$ is, however, slightly more
involved and require some additional modelling. We discuss the
procedure that we have adopted to estimate these quantities in the
following sections.

\subsection{Estimate of the neutral gas fraction $f_{{\rm neutr},i}$}
\label{sec:neutrgas}

\FM{The gas cooling module in \arepo\ determines $f_{{\rm neutr},i}$ for each 
cell. This quantity is computed by modelling an ionization network for a mixture 
of hydrogen and helium \citep{Katz1992, Katz1996}. Two body processes (such as 
collisional excitation, collisional  ionization, recombination, dielectric 
recombination and free--free emission) and Compton cooling off the cosmic 
microwave background \citep{Ikeuchi1986} are considered. In addition to these, 
the energy input due to a time-varying photoionizing UV background 
\citep{Faucher-Giguere2009} is also taken into account. For a complete 
description of the gas cooling module and metal cooling treatment used in Auriga 
we refer the reader to \citet{Vogelsberger2013}.}

\FM{In principle, this information could be directly used in 
equation (\ref{eq:HImass}).} However, in the galaxy formation model used in the 
simulations, gas above a predetermined density threshold, $n_{\rm 
th}\sim0.13\,{\rm cm^{-3}}$, is eligible for star formation. This star-forming 
gas, in turn, is modelled by an effective equation of state describing a 
simplified ISM model comprising hot ($T\sim 10^7\,\K$) and cold ($T\sim 
10^3\,\K$) gas phases in pressure equilibrium \citep[see][]{Springel2003}. 
Because the thermodynamic state of the gas is dictated by the effective equation 
of state, the value $f_{{\rm neutr},i}$ estimated by the code for the 
star-forming gas is unreliable for our purposes.

As stated above, the cold gas phase, which is one of the constituents
of our ISM model, has a temperature around $10^3\,\K$ and as such we
expect that the gas comprising it is essentially neutral. Therefore,
it is natural to assume that for star-forming gas, $f_{{\rm neutr},i}$
is simply the mass fraction of the cold gas phase of our ISM, which
can be easily determined as \citep[see][]{Springel2003}
\begin{equation}
x = \frac{u_h - u}{u_h - u_c},
\label{eq:xfactor}
\end{equation}
where $u = P/[(\gamma - 1)\rho]$ is the gas thermal energy per unit
mass (set by the effective equation of state), and $u_c$ and $u_h$ are
the gas thermal energy of the cold and the hot ISM phase,
respectively. In what follows we will thus use for $f_{{\rm neutr},i}$
the value estimated from equation (\ref{eq:xfactor}) if the gas is
star-forming, or otherwise the value directly computed by the \arepo\
cooling module for gas densities below the star formation threshold
$n_{\rm th}$.

\subsection{Treatment of the molecular gas}\label{sec:molgas}

Once $f_{{\rm neutr},i}$ has been determined, the contribution of the
atomic hydrogen must be separated from the molecular phase. The galaxy
formation physics modules currently employed in \arepo\ do not directly account
for the mechanisms responsible for the creation and destruction of
molecular hydrogen. Therefore, we adopt two phenomenological
methods to achieve this objective, also with the goal to investigate
systematic differences arising from the specific treatment of
molecular hydrogen in our analysis (see Section~\ref{sec:results} for
more details).

The first method is purely empirical and is based on a set of
observations of nearby galaxies \citep[][hereafter
\citetalias{Leroy2008}]{Leroy2008}. As originally proposed by
\citet{Blitz2006}, it consists of fitting the ratio $R_{\rm mol}$
between the column density of molecular over atomic hydrogen with the
functional form
\begin{equation}
 R_{\rm mol} = \left(\frac{P}{P_0}\right)^\alpha,
 \label{eq:Blitz}
\end{equation}
where $P$ is the gas mid-plane pressure and $P_0$ and $\alpha$ are
free parameters.  In our case, \FM{we take $R_{\rm mol}$ in
equation (\ref{eq:Blitz}) as the ratio between the molecular and atomic
hydrogen volumetric densities and $P$ as the gas pressure\footnote{We consider only the
  partial pressure of the cold gas phase\FM{, computed by multiplying the total gas pressure by the 
  neutral gas fraction $x$ in equation (\ref{eq:xfactor}), for star-forming gas.}}
  in any given cell.}
  The molecular fraction can then be easily obtained as
\begin{equation}
 f_{{\rm mol},i} = \frac{R_{{\rm mol},i}}{R_{{\rm mol},i} + 1},
\end{equation}
once the value of $P_0$ and $\alpha$ have been fixed. Following
\citetalias{Leroy2008} we choose
$P_0 = 1.7\times10^4\,{\rm \K\,cm^{-3}}$ and $\alpha = 0.8$. We
explicitly checked that using the parametrization proposed by
\citet{Blitz2006} does not alter our findings significantly.

The second method is more theoretically motivated and is based on the
work by \citet[][hereafter \citetalias{Gnedin2011}]{Gnedin2011}.
Following \citet[][Appendix A1]{Lagos2015} we define
\begin{equation}
 f_{{\rm mol},i} = \left(1 + \frac{\Sigma_c}{\Sigma_{{\rm neutr},i}}\right)^2,
 \label{eq:Lagos}
\end{equation}
where $\Sigma_{{\rm neutr},i}$ is the neutral hydrogen column density and $\Sigma_c$ is a critical 
column density given by
\begin{equation}
 \Sigma_{c}=20\,{\rm M}_{\odot}\,{\rm pc}^{-2} \frac{\Lambda(D_{\rm MW}, G^{'}_0)^{4/7}}{D_{\rm MW}\sqrt{1+G^{'}_0\,D^2_{\rm MW}}}.
 \label{eq:sigmac}
\end{equation}
In the previous expression, $D_{\rm MW}$ is the dust-to-gas mass ratio
normalized to the Milky Way value and estimated as
$D_{\rm MW} \equiv Z_i/Z_{\odot}$, where $Z_i$ is the metallicity of
the gas cell, $Z_{\odot}=0.0127$, $G^{'}_0$ is the
interstellar radiation field in units of the \citet{Habing1968}
radiation field, and $\Lambda(D_{\rm MW}, G^{'}_0)$ is a fitting
function depending on the latter two variables (see
\citetalias{Gnedin2011}; \citealt{Lagos2015}, for more details). As in
\citet{Lagos2015}, $G^{'}_0$ is expressed in terms of the star
formation rate (SFR) density in the solar neighbourhood as
$G^{'}_0=\Sigma_{\rm SFR}/\Sigma_{\rm SFR,0}$, where
$\Sigma_{\rm SFR,0}=10^{-3}\,\rm M_{\odot}\, \rm yr^{-1}\,\rm
kpc^{-2}$
\citep{Bonatto2011}.  Column (surface) densities are computed by
multiplying the relevant volumetric quantities by the Jeans length
\citep[see][]{Schaye2001}
\begin{equation}
\lambda_{\rm J}=\frac{c_{\rm s}}{\sqrt{G\,\rho}},
\label{eq:Jeans}
\end{equation}
where $c_{\rm s}$ is the sound speed, $G$ the gravitational constant,
and $\rho$ the total gas density. $\lambda_{\rm J}$ is computed
individually for every cell. Finally, to protect against numerical
divergences, a metallicity floor equal to $10^{-8}\, Z_{\odot}$ is
imposed.

\begin{figure*}
\centering
\includegraphics[width=0.19\textwidth]{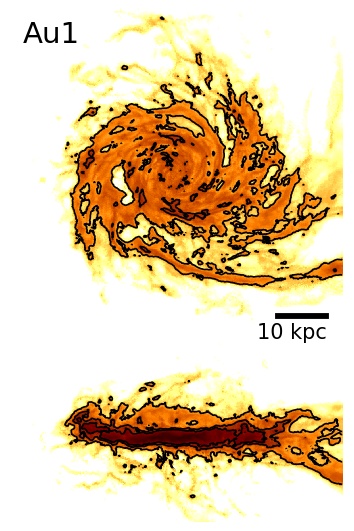}
\includegraphics[width=0.19\textwidth]{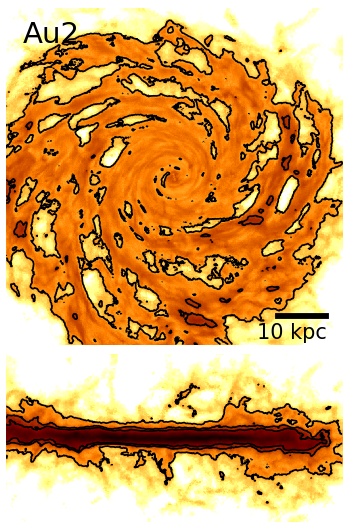}
\includegraphics[width=0.19\textwidth]{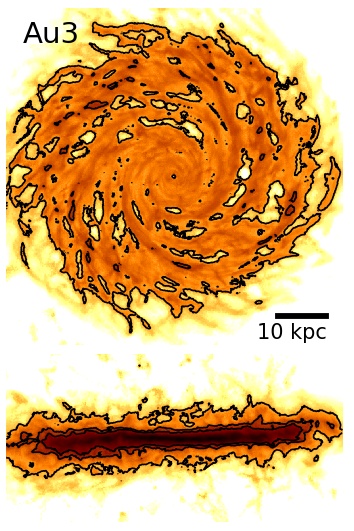}
\includegraphics[width=0.19\textwidth]{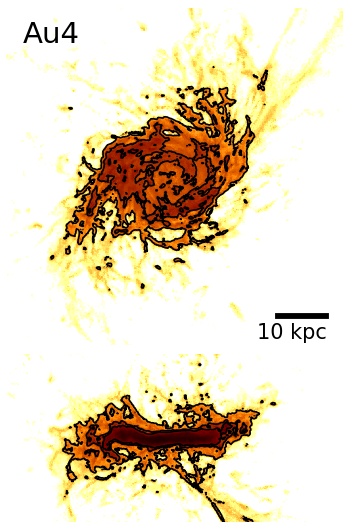}
\includegraphics[width=0.19\textwidth]{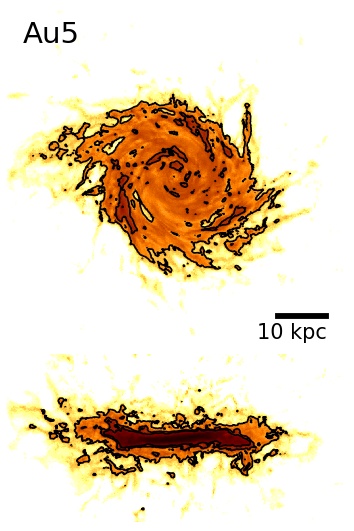}
\includegraphics[width=0.19\textwidth]{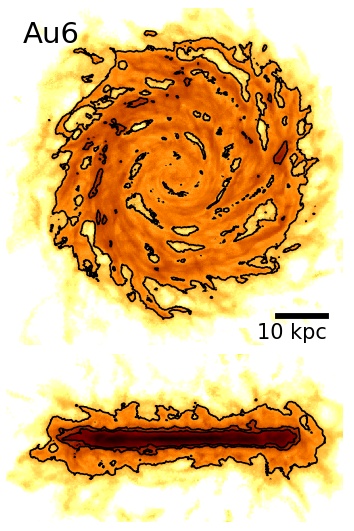}
\includegraphics[width=0.19\textwidth]{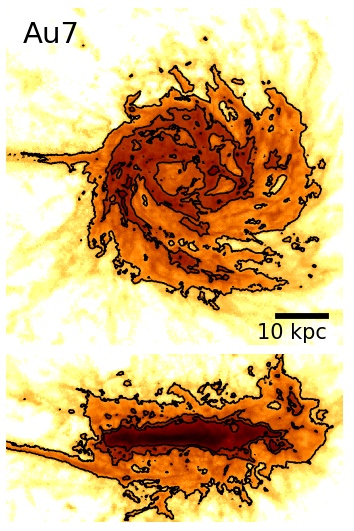}
\includegraphics[width=0.19\textwidth]{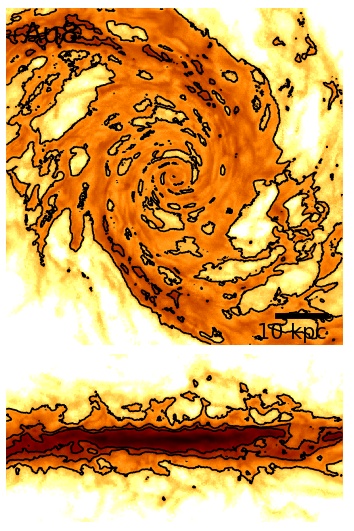}
\includegraphics[width=0.19\textwidth]{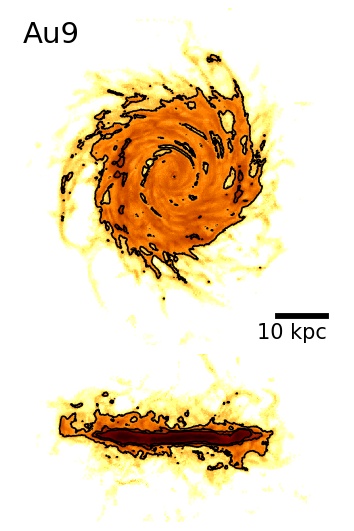}
\includegraphics[width=0.19\textwidth]{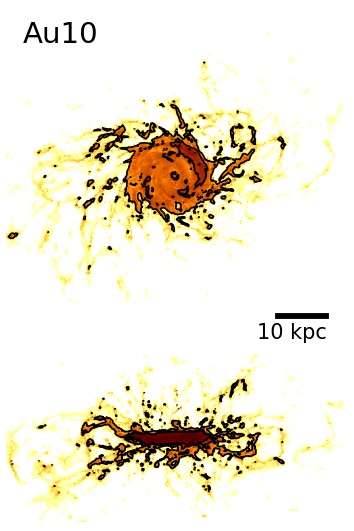}
\includegraphics[width=0.19\textwidth]{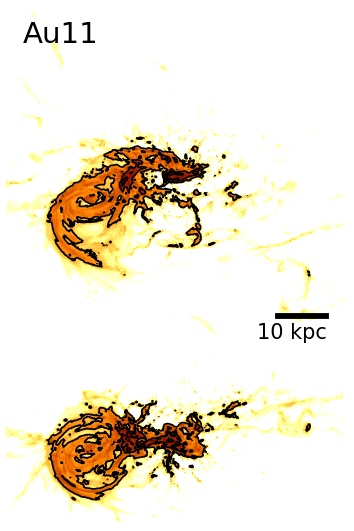}
\includegraphics[width=0.19\textwidth]{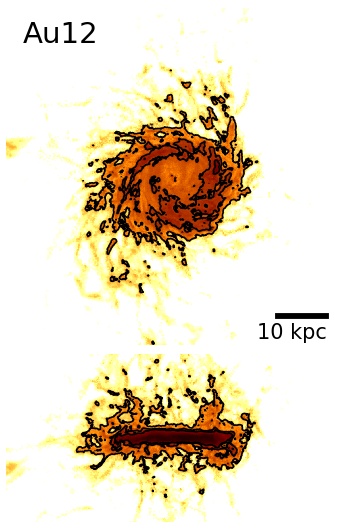}
\includegraphics[width=0.19\textwidth]{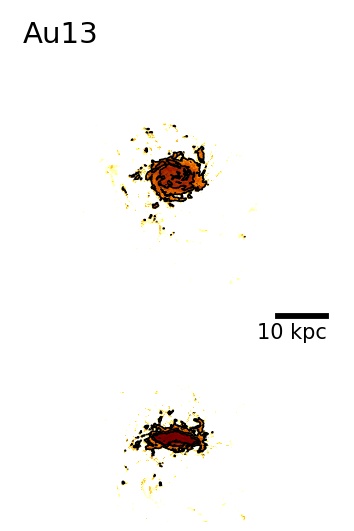}
\includegraphics[width=0.19\textwidth]{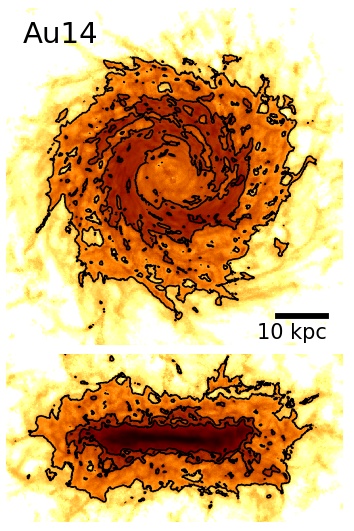}
\includegraphics[width=0.19\textwidth]{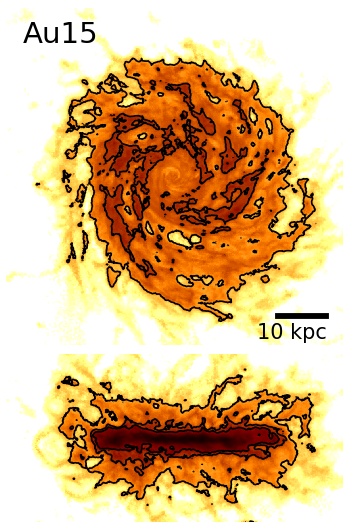}
\caption{\hi\ projections for haloes Au~1 to Au~15 obtained by using
  the \citetalias{Leroy2008} empirical relation to account for
  molecular gas. Each halo is shown face-on (top) and edge-on
  (bottom). The size of the projection region is 90 kpc on a side for
  face-on projections and $90\times 45$ kpc for the edge-on views.}
\label{fig:HIimages1}
\end{figure*}

Fig.~\ref{fig:HIprojections} illustrates the different steps of the
procedure that we have just described when applied to halo Au 5. In
particular, the panels from left to right can be thought of as the result
of progressively applying the correcting factors (from the rightmost
to the leftmost) used to determine the \hi\ content listed in
equation (\ref{eq:HImass}) to the total gas cell mass. They show,
respectively, the projection both face on (top row) and edge on
(bottom row) of the total H density (including ionized gas), the
neutral H density (\hi\ + H$_2$) and the \hi\ density obtained
following the \citetalias{Leroy2008} and \citetalias{Gnedin2011}
prescriptions.

\begin{figure*}
\centering
\includegraphics[width=0.19\textwidth]{fig3a}
\includegraphics[width=0.19\textwidth]{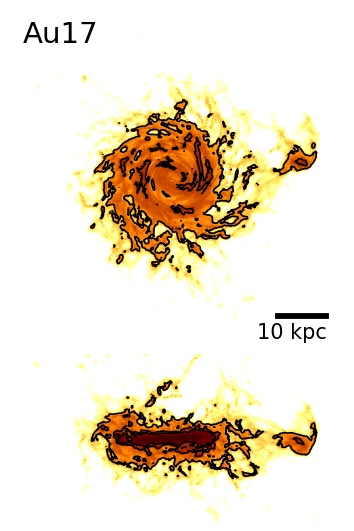}
\includegraphics[width=0.19\textwidth]{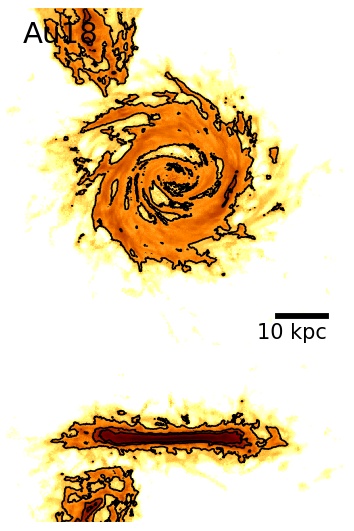}
\includegraphics[width=0.19\textwidth]{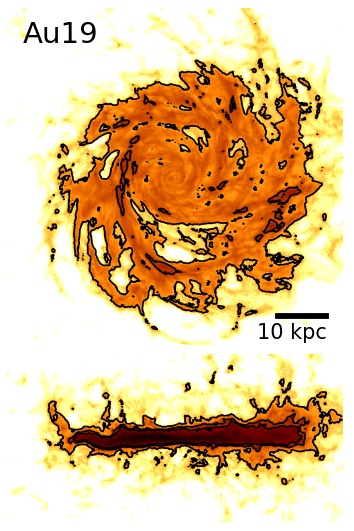}
\includegraphics[width=0.19\textwidth]{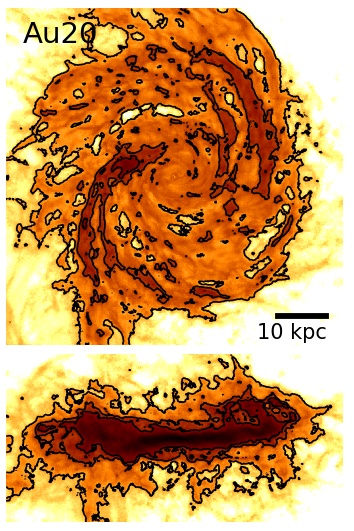}
\includegraphics[width=0.19\textwidth]{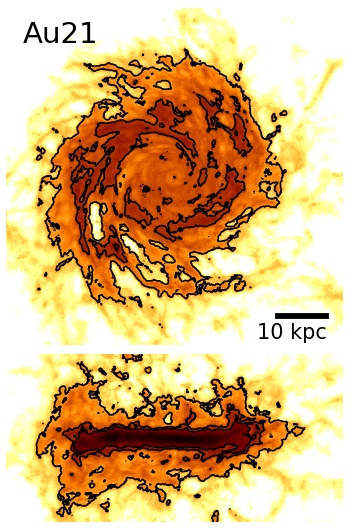}
\includegraphics[width=0.19\textwidth]{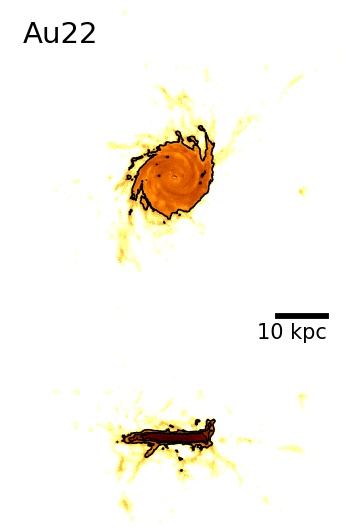}
\includegraphics[width=0.19\textwidth]{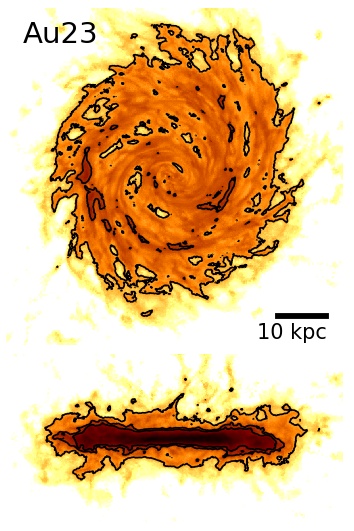}
\includegraphics[width=0.19\textwidth]{fig3i}
\includegraphics[width=0.19\textwidth]{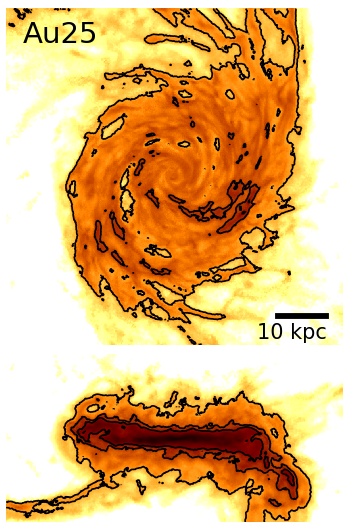}
\includegraphics[width=0.19\textwidth]{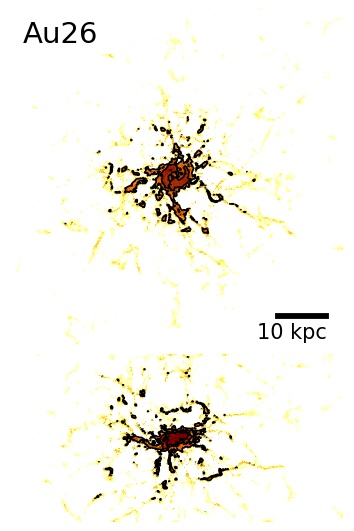}
\includegraphics[width=0.19\textwidth]{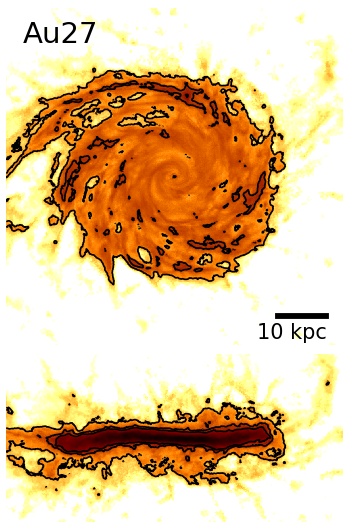}
\includegraphics[width=0.19\textwidth]{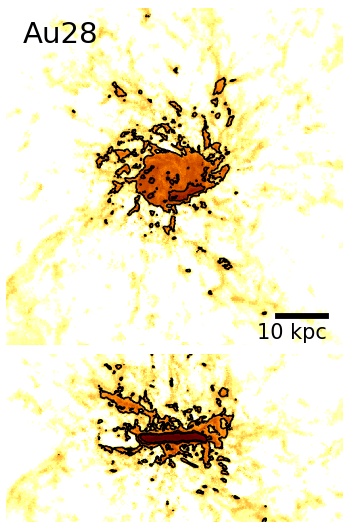}
\includegraphics[width=0.19\textwidth]{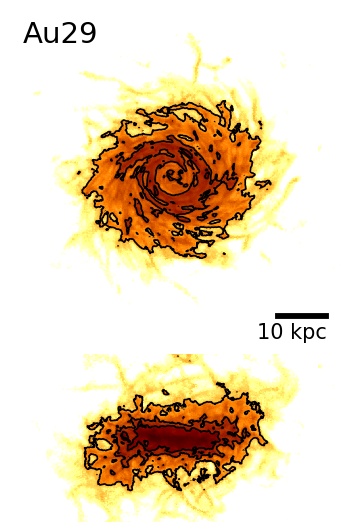}
\includegraphics[width=0.19\textwidth]{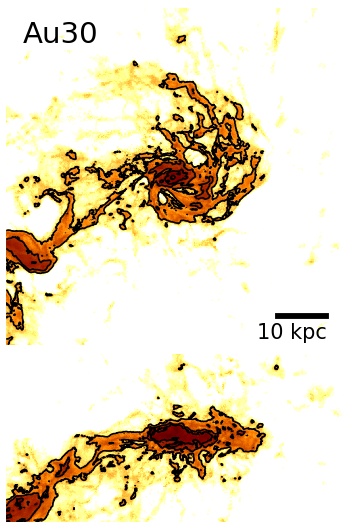}
\caption{\hi\ projections for haloes Au~16 to Au~30 obtained by using
  the \citetalias{Leroy2008} empirical relation to account for
  molecular gas. Each halo is shown face on and edge on. The size of
  the projection region is 90 kpc on a side for face-on projections
  and $90\times 45$ kpc for the edge-on views.}
\label{fig:HIimages2}
\end{figure*}

\begin{figure*}
\centering
\includegraphics[width=0.19\textwidth]{fig4a}
\includegraphics[width=0.19\textwidth]{fig4b}
\includegraphics[width=0.19\textwidth]{fig4c}
\includegraphics[width=0.19\textwidth]{fig4d}\\
\includegraphics[width=0.19\textwidth]{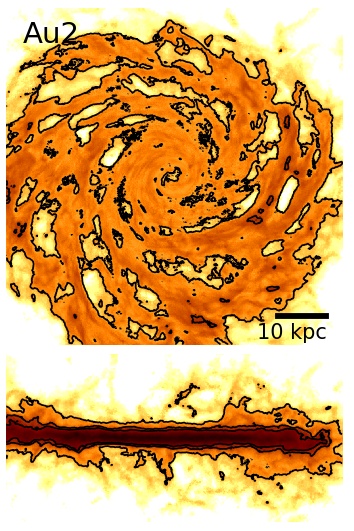}
\includegraphics[width=0.19\textwidth]{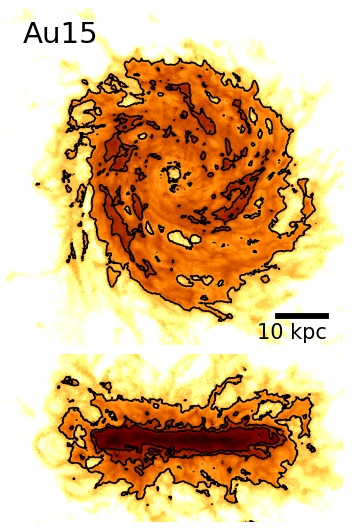}
\includegraphics[width=0.19\textwidth]{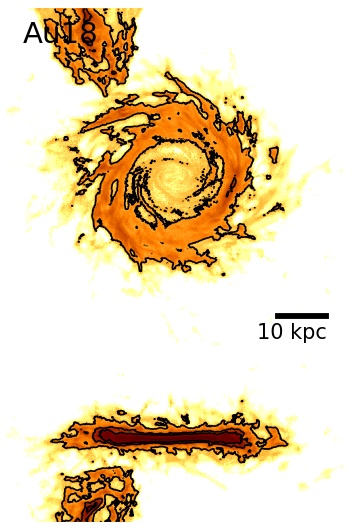}
\includegraphics[width=0.19\textwidth]{fig4h}
\caption{Comparison of the \hi\ disc morphologies obtained by using
  the \citetalias{Leroy2008} (top row) and \citetalias{Gnedin2011} (bottom row) 
  prescriptions to account for molecular gas for a selection of four Auriga haloes. 
  Each halo is shown face-on and edge-on, with a size of the projection region of 90 kpc
  on a side for face-on projections and  $90\times 45$ kpc for edge-on projections.}
\label{fig:HIimages3}
\end{figure*}

It is readily apparent in these projections that the disc of neutral
hydrogen in the Au 5 halo is surrounded by a much more extended layer
of ionized gas. Removing the ionized layer leaves a gaseous disc that
has about the same extent and external shape whether or not molecular
hydrogen is taken into account. This is consistent with the picture
that molecules form in the highest density regions (i.e.~in the
central regions of the simulated discs), and their importance to the
mass budget contribution in the outskirts is negligible. The largest
difference between the total (\hi + H$_2$) neutral hydrogen and \hi\
column density is certainly in the centre (the innermost
$\sim 10\,\kpc$), where the projected density drops by a factor of
$\sim 10$.  Although there is a very good resemblance in the main
morphological features of the \hi\ projection regardless of the method
used to take into account the molecular gas, some minor differences
are visible, and in particular the \citetalias{Gnedin2011} method
yields consistently lower \hi\ column densities in the central regions
of the simulated disc. As we will discuss more extensively later on,
this appears to be a general trend and it is not limited to the
particular halo that we have presented here.

\section{Results}\label{sec:results}

\subsection{\hi\ morphologies}\label{sec:morphologies}
We start our analysis by presenting in Figs.~\ref{fig:HIimages1} and
\ref{fig:HIimages2} \hi\ density projections for the set of 30 Auriga
galaxies at redshift $z=0$ obtained with the \citetalias{Leroy2008}
method to estimate the molecular gas fraction.
Both face-on (top panels) and edge-on projections
(bottom panels) are presented. The gaseous discs are aligned by
computing the spin axis of the dense, star-forming gas within 10\% of
the virial radius of each object. Each panel has the same physical
size ($90\times 90$ kpc for face-on projections, and $90\times 45$ kpc
for the edge-on ones, with a depth of projection of 90 kpc in all
cases) to facilitate the size comparison among different objects, and
is centred on the potential minimum of each simulated galaxy. The
colour mapping, encoding column densities in the range
$5\times10^{18}-10^{22}\,\cm^{-2}$, is also kept the same in all panels
for the same reason. Contours are placed at \hi\ surface densities of
$10$, $5$, and $1\,{\rm M}_{\odot}\,\pc^{-2}$ from the innermost to
the outermost. \FM{We remind the reader that $1\,{\rm M}_{\odot}\,\pc^{-2} 
\simeq 1.249\times 10^{20}\,\cm^{-2}$ for hydrogen.}

In the majority of the systems, the \hi\ gas is distributed in an extended disc 
(i.e.~is more extended than the stellar component), which in some cases can 
reach galactocentric distances up to $\sim 60\,\kpc$ (e.g. Au 2, Au 16, Au 24).  
In the vertical direction, the thickness of the \hi\ layer is very diverse. We 
have examples of relatively thin gaseous discs (e.g.~Au 2, Au 9, Au 10, Au 18, 
Au 19, Au 23, Au 24) while others feature a very extended vertical distribution 
(Au 7, Au 14, Au 15, Au 21, Au 25, Au 29) that can reach tens of kpc outside the 
mid-plane. In the latter case, these vertically extended layers are often found 
in conjunction with a ring of more dense \hi\ material or gas that is tracking 
spiral-like features in the face-on projection. These features are much less 
frequent (and less prominent) for thinner \hi\ discs. 
\FM{The vertical thickness can be of the order of $\sim1.5-2\,\kpc$ for thin 
discs, while reaching $\sim 5\,\kpc$ for more extended structures (see 
Table~\ref{tab:hiproperties})}. In Section~\ref{sec:vertical} 
we discuss the \FM{calculation of the} disc vertical thickness in more detail 
and relate it to the galactic fountain flows generated by stellar feedback.

A comparison between the \hi\ morphologies obtained with the two
prescriptions to determine the molecular gas fraction is presented in
Fig.~\ref{fig:HIimages3} (top row showing the \citetalias{Leroy2008},
method and bottom row the \citetalias{Gnedin2011} method,
respectively) for a few selected examples.  In agreement with the
results discussed in Section \ref{sec:molgas}, the morphologies of the
\hi\ discs remain relatively unchanged. For example, their vertical
and radial extent is only marginally affected by how the contribution
of H$_2$ is calculated (see also Table~\ref{tab:hiproperties} and
Section \ref{sec:massdiff} for a more quantitative analysis). The main
difference is again in the central regions of the disc, where the
\citetalias{Gnedin2011} approach predicts a lower \hi\ fraction than
the \citetalias{Leroy2008} method. This is illustrated by the decrease
of the \hi\ surface density, which sometimes, like in Au 18 and Au 24,
can be very pronounced and extend over large ($\sim 10\,\kpc$) scales.

It is also informative to compare our \hi\ morphology with that obtained in 
other cosmological simulations of galaxy formation, and in particular with the 
recent \hi\ study of \citetalias{Bahe2016} based on the \eagle\ simulations 
\citep{Schaye2015}.  An interesting aspect of this comparison is that similar 
methods were used to estimate the \hi\ fraction in both the {\sc eagle} and 
Auriga simulations, whereas the two simulation sets differ in the implementation 
of the ISM physics \FM{-- although they both use an effective equation of state 
for cold and dense gas \citep[see][respectively]{DallaVecchia2012,Springel2003} 
--} and stellar feedback, \FM{as well as in the achieved resolution, which is 
about a factor of 36 better in mass for the Auriga runs}. The main difference 
lies in the treatment of stellar feedback, which is local and more violent in 
{\sc eagle} compared to the Auriga model. The latter employs a stochastic 
spawning of wind `particles' that effectively gives rise to a non-local 
injection of supernova momentum \citep[see][for details]{Vogelsberger2013}. This 
has a significant impact on the \hi\ morphologies. In the {\sc eagle} 
simulations, unrealistically large holes of \FM{tens of kpc of size} are clearly 
visible in the neutral gas distribution \FM{of some galaxies}. This affects not only the morphologies 
of the \hi\ discs, but also the radial gas distribution. Our simulations do not 
seem to suffer from the same problem, although gaps in the \hi\ distribution can 
sometimes be seen, especially for the more extended \hi\ discs (e.g. Au 
2, Au 16, Au 24). \FM{To conclude, it is conceivable that some of the 
adverse effects on the \hi\ distribution due to the nature of the stellar 
feedback used in the \eagle\ simulation might be mitigated at higher resolution. 
Indeed, resolution is expected to have some impact on the efficiency 
\citep[see][]{DallaVecchia2012} and the `granularity' (i.e.~ the number of gas 
particles heated per feedback event) of the stellar feedback in \eagle, making 
it less violent and helping reducing the unphysically large sizes of the \hi\ 
holes.}

\begin{table*}
\begin{tabular}{ccccccccc}
\hline
     Run   &    $R_{\rm HI}^{\rm L}$ & $M_{\rm HI}^{\rm L}$ & $\Sigma_{\rm HI}^{\rm L}$ & $h_{\rm HI}^{\rm L}$ & $R_{\rm HI}^{\rm GK}$ & $M_{\rm HI}^{\rm GK}$ & $\Sigma_{\rm HI}^{\rm GK}$ & $h_{\rm HI}^{\rm GK}$\\
           &    $[\kpc]$ & $[10^9\,{\rm M_{\odot}}]$ & $[{\rm M_{\odot}}\,\pc^{-2}]$ & $[\kpc]$ & $[\kpc]$ & $[10^9\,{\rm M_{\odot}}]$ & $[{\rm M_{\odot}}\,\pc^{-2}]$ & $[\kpc]$ \\
\hline
\multicolumn{9}{c}{Fiducial resolution (level 4)} \\
\hline
     Au 1  &    29.6 &    8.0 &  2.9 &  1.9 &   29.9 &      8.7 &      3.1  &  2.4\\
     Au 2  &    49.9 &   22.3 &  2.8 &  1.7 &   50.0 &     22.0 &      2.8  &  1.8\\
     Au 3  &    42.0 &   18.2 &  3.3 &  1.7 &   42.2 &     17.0 &      3.0  &  2.3\\ 
     Au 4  &    22.7 &    7.3 &  4.5 &  2.7 &   23.5 &      7.6 &      4.4  &  4.9\\
     Au 5  &    24.4 &    7.2 &  3.8 &  1.9 &   24.7 &      7.0 &      3.7  &  2.9\\ 
     Au 6  &    38.0 &   15.0 &  3.3 &  1.6 &   38.2 &     14.8 &      3.2  &  1.9\\
     Au 7  &    35.4 &   17.9 &  4.6 &  4.3 &   35.5 &     19.1 &      4.8  &  5.8\\
     Au 8  &    54.8 &   22.1 &  2.3 &  1.8 &   55.1 &     22.9 &      2.4  &  1.9\\
     Au 9  &    24.5 &    6.3 &  3.3 &  1.3 &   24.7 &      6.3 &      3.3  &  1.8\\
     Au 10 &    13.2 &    2.2 &  3.9 &  1.7 &   13.6 &      2.4 &      4.1  &  4.1\\
     Au 11 &    15.1 &    1.4 &  1.9 &  3.5 &   15.1 &      1.7 &      2.3  &  4.5\\
     Au 12 &    19.8 &    6.6 &  5.3 &  2.2 &   20.2 &      7.0 &      5.5  &  4.8\\
     Au 13 &     7.1 &    0.8 &  5.0 &  1.7 &    7.3 &      1.1 &      6.8  &  2.5\\
     Au 14 &    34.6 &   21.4 &  5.7 &  4.5 &   34.8 &     21.2 &      5.6  &  6.2\\
     Au 15 &    34.0 &   17.2 &  4.7 &  3.1 &   34.1 &     17.1 &      4.7  &  4.7\\
     Au 16 &    52.2 &   29.2 &  3.4 &  2.0 &   52.3 &     29.2 &      3.4  &  2.4\\
     Au 17 &    20.3 &    4.4 &  3.4 &  2.2 &   20.5 &      4.7 &      3.5  &  3.4\\
     Au 18 &    25.0 &    5.0 &  2.5 &  1.2 &   25.1 &      4.6 &      2.3  &  1.5\\
     Au 19 &    36.2 &   12.9 &  3.1 &  1.5 &   36.3 &     12.7 &      3.1  &  1.9\\
     Au 20 &    46.9 &   29.3 &  4.3 &  3.7 &   47.1 &     30.8 &      4.4  &  4.3\\
     Au 21 &    34.8 &   17.7 &  4.7 &  2.6 &   34.9 &     17.0 &      4.4  &  4.5\\
     Au 22 &    11.2 &    1.4 &  3.6 &  1.0 &   11.3 &      1.4 &      3.4  &  1.5\\
     Au 23 &    34.8 &   14.5 &  3.8 &  1.7 &   35.0 &     14.2 &      3.7  &  2.1\\
     Au 24 &    43.9 &   18.3 &  3.0 &  1.3 &   44.0 &     18.0 &      2.9  &  1.7\\
     Au 25 &    37.8 &   15.6 &  3.5 &  2.5 &   37.9 &     15.6 &      3.5  &  3.0\\
     Au 26 &     5.6 &    0.5 &  5.0 &  3.7 &    5.9 &      0.7 &      6.7  & 10.4\\
     Au 27 &    32.0 &   13.7 &  4.3 &  1.9 &   32.2 &     13.0 &      4.0  &  2.7\\
     Au 28 &    14.7 &    1.9 &  2.9 &  4.8 &   14.9 &      2.1 &      3.0  &  9.2\\
     Au 29 &    22.6 &    8.3 &  5.2 &  4.5 &   22.7 &      7.9 &      4.9  &  5.3\\
     Au 30 &    18.0 &    2.5 &  2.4 &  2.7 &   19.1 &      2.8 &      2.4  &  3.8\\
\hline
\multicolumn{9}{c}{Low resolution (level 5)} \\
\hline
     Au 16 &    61.1 &   39.1 &  3.3 &  3.4 &   61.4 &     40.8 &      3.5  &  3.4\\
     Au 24 &    30.5 &   13.1 &  4.5 &  2.0 &   30.7 &     13.5 &      4.6  &  2.3\\
\hline
\multicolumn{9}{c}{High resolution (level 3)} \\
\hline
     Au 16 &    57.9 &   27.2 &  2.6 &  1.6 &   58.2 &     26.6 &      2.5  &  2.3\\
     Au 24 &    43.1 &   17.8 &  3.0 &  1.8 &   43.3 &     17.0 &      2.9  &  2.4\\
\hline
\end{tabular}
\caption{Properties of \hi\ discs in the Auriga simulations. The columns list (from left to right): simulation name; \hi\ radius, \hi\ mass, 
  average \hi\ surface density and \hi\ vertical thickness (\citetalias{Leroy2008} method); \hi\ radius, 
  \hi\ mass, average \hi\ surface density and \hi\ vertical thickness (\citetalias{Gnedin2011} method).
} 
\label{tab:hiproperties}
\end{table*}

\subsection{\hi\ radial profiles}\label{sec:profiles}

\begin{figure*}
\centering
\includegraphics[width=0.77\textwidth]{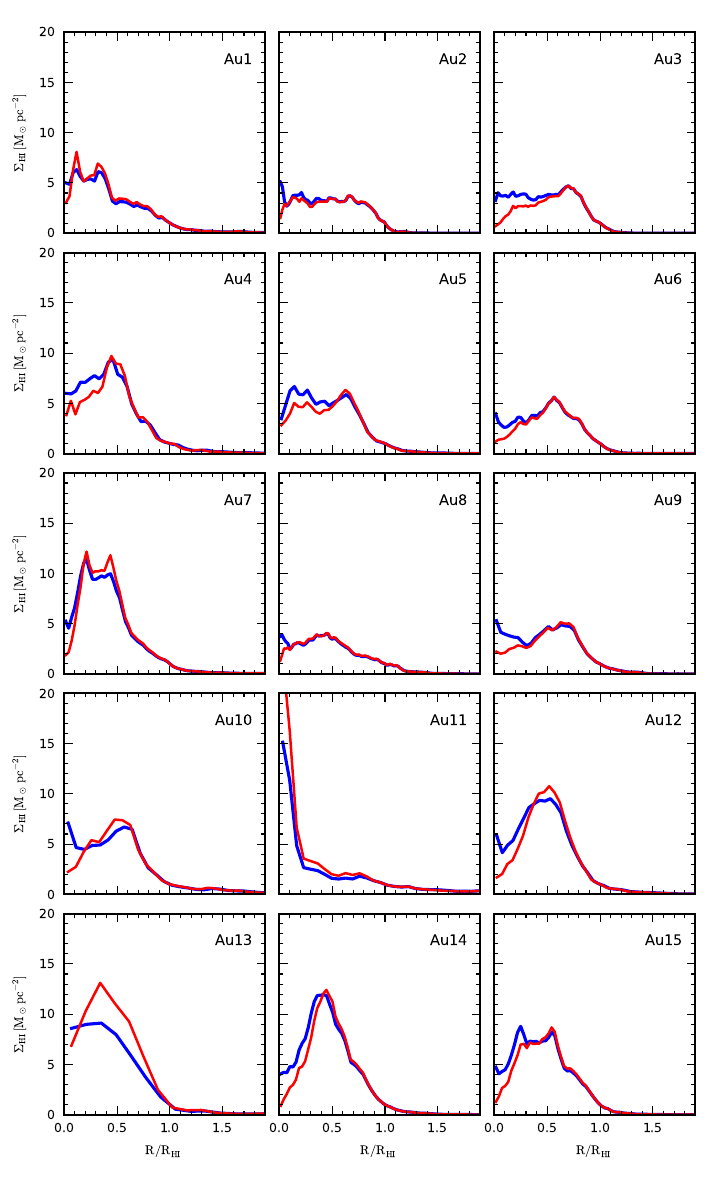}
\caption{\hi\ surface density profiles as a function of the
  galactocentric radius (in units of the \hi\ radius of each galaxy,
  see Table~\ref{tab:hiproperties}) for haloes Au~1 to Au~15. The
  blue and red lines indicate the profiles obtained using the
  prescription from \citetalias{Leroy2008} or \citetalias{Gnedin2011}
  for removing molecular gas, respectively. }
\label{fig:HIprofiles1}
\end{figure*}

\begin{figure*}
\centering
\includegraphics[width=0.77\textwidth]{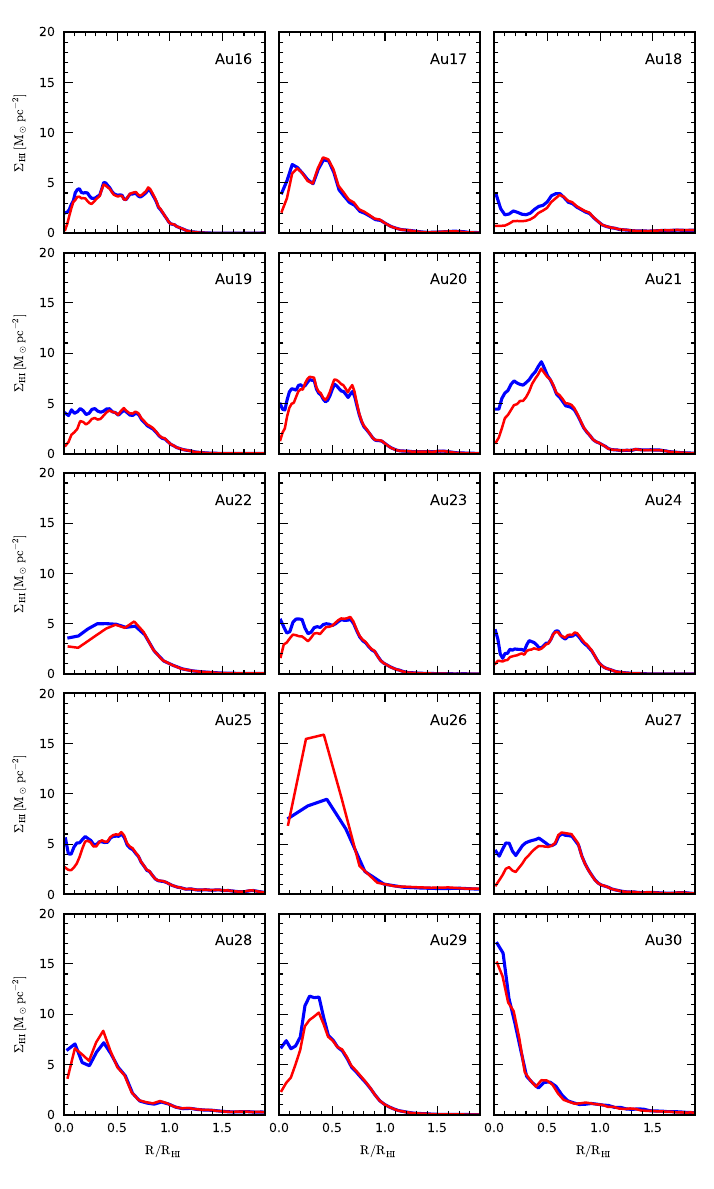}
\caption{As in Fig.~\ref{fig:HIprofiles1}, but for haloes Au~16
  to Au~30.}
\label{fig:HIprofiles2}
\end{figure*}

Figs \ref{fig:HIprofiles1} and \ref{fig:HIprofiles2} show azimuthally
averaged \hi\ surface density profiles as a function of the
galactocentric radius for the 30 objects in the Auriga
simulations. The galactocentric radii have all been expressed in terms
of the azimuthally averaged \hi\ radius of each object, defined as
$\Sigma_{\rm HI}(R_{\rm HI}) = 1\,{\rm M}_{\odot}\,{\rm pc^{-2}}$
\citep[see, e.g.,][and Table~\ref{tab:hiproperties} for their
values]{Broeils1997}.  The profiles have been obtained by turning the
\hi\ disc face on (using again a direction defined by the angular
momentum of the cold star-forming gas) and considering the gas cells
in a \FM{rectangular box centred on the halo potential minimum and extending for 
$120$ kpc perpendicular to the disc and for $240\times240$ kpc in the other
directions}. The gas cells in this region have been binned in 120
concentric annuli of 1 kpc width, and the surface density of each
annulus has been computed by calculating the total \hi\ mass of the
annulus divided by its surface area. The figure shows the profiles for
both the \citetalias{Leroy2008} (blue) and \citetalias{Gnedin2011}
(red) method of computing the molecular fraction.

Although some variability is present between the different profiles, a
general trend can be identified. Starting from the centre of the
profile, the \hi\ surface density increases up to a point where it
reaches a plateau of approximately constant surface density. After
this plateau, which in some cases (e.g. Au 2, Au 16, Au 20) can extend
up to 10 kpc, $\Sigma_{\rm HI}$ drops quite rapidly, and very little
\hi\ is left past $R_{\rm HI}$. The density plateau is not always
present. Some profiles (e.g. Au 6, Au 12, Au 13, Au 14, Au 21, Au 26)
instead show an increase of $\Sigma_{\rm HI}$ up to a maximum value
after which the profiles immediately decline. Other profiles (e.g. Au
11, Au 30) show a monotonic decrease of the \hi\ surface density at
all radii, but we note that this could be due to interactions with
close, massive companions that they are currently experiencing.

It is also interesting to compare the difference in the \hi\ radial profiles due 
to the two different methods that we have used to take into account the 
contribution of the molecular gas to the total neutral H density. In general, it 
can be seen that the two methods agree very well at large radii (i.e.~$R\sim 
R_{\rm HI}$), where the contribution of molecular hydrogen to the total budget 
is negligible. The agreement still remains remarkable at intermediate radii, 
where the surface density plateau or the maximum value of $\Sigma_{\rm HI}$ are 
reached, but close to the centre, the \citetalias{Gnedin2011} method generally 
results in more depressed profiles (but see Au 11, Au 13, Au 26 for the opposite 
trend). In Section~\ref{sec:massdiam}, we discuss in more detail what these 
differences imply for the global properties (such as the total \hi\ mass) and 
scaling relations of the simulated gas discs.

\begin{figure}
\centering
\includegraphics[width=0.43\textwidth]{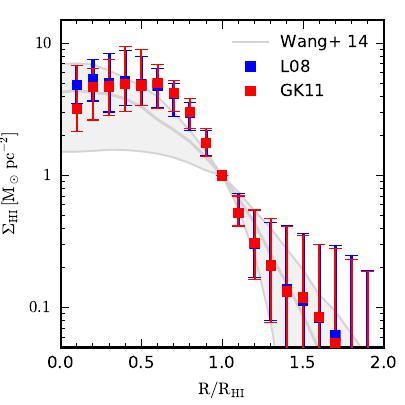}
\caption{Median \hi\ profiles for the 30 Auriga haloes derived with
  the \citetalias{Leroy2008} (blue symbols) and
  \citetalias{Gnedin2011} (red symbols) prescriptions for the
  treatment of molecular hydrogen. For comparison, the median \hi\
  profile for the Bluedisk sample \citep[][grey lines]{Wang2014} is
  shown. Error bars and the shaded grey region indicate the area
  between the 16-th and 84-th percentiles of each radial bin (of size
  $0.1\,R_{\rm HI}$) of the three distributions.  }
\label{fig:HIprofilesmed}
\end{figure}

In Fig.~\ref{fig:HIprofilesmed} we show the median \hi\ radial
profiles for the 30 Auriga haloes computed with the
\citetalias{Leroy2008} (blue symbols with error bars) and the
\citetalias{Gnedin2011} (red symbols with error bars) prescriptions
for estimating the molecular hydrogen fraction. The median values have
been obtained by interpolating the profiles presented in
Figs.~\ref{fig:HIprofiles1} and \ref{fig:HIprofiles2} as a function of
the galactocentric distance (normalized to $R_{\rm HI}$) on to a
uniformly spaced radial grid of width $0.1\times R/R_{\rm HI}$. Error
bars indicate the region between the 16-th and 84-th percentiles of
each radial bin of the simulated \hi\ profiles. The simulation results
are compared with the median profile (obtained with the same
procedure) of the Bluedisk sample \citep[][grey solid
line]{Wang2014}. The shaded grey area indicates again the region
between the 16-th and 84-th percentiles in each radial bin of the
observed Bluedisk radial profiles.

A comparison between the median simulated profiles reveals that they are very 
similar at all radii, and that they start to differ only in the innermost 
regions where the \citetalias{Gnedin2011} profile starts to show a decrease in 
the \hi\ surface density, which is however less marked than the one observed in 
the individual profiles. Compared to the Bluedisk sample, both the 
\citetalias{Leroy2008} and the \citetalias{Gnedin2011} median profiles are 
largely consistent with observations although some discrepancies are present. In 
particular, the simulated profiles overpredict the values of the \hi\ 
surface density in the radial range $0.4 \lsim R/R_{\rm HI} \lsim 0.8$.

It is also instructive to compare our results with those found in the \eagle\ 
simulations \citepalias[][]{Bahe2016}, in which the more local and violent 
nature of the stellar feedback results in $\sim\,10$ kpc-scale holes in the 
(neutral) gas distribution. 
\FM{We note that \hi\ holes have been observed in 
nearby disc galaxies, although they have more moderate sizes
($\sim\,1$ kpc), and are indeed interpreted as an effect of stellar feedback \citep[][and references therein]{Boomsma2008}.} 
The presence of these large holes impacts the \hi\ radial profiles of the \eagle\ 
galaxies, with the result that in the central parts the median \hi\ profile is 
underestimated with respect to the observations. The Auriga galaxies, thanks to 
a more distributed stellar feedback, do not suffer from this problem. 
In fact, they tend to overpredict the \hi\ content at intermediate radii. 
Moreover, the discrepancies in the radial profiles
(but also other \hi\ related quantities; see the following sections) due to a
different choice of the prescription used to model the contribution of
molecular hydrogen are negligible in the Auriga case (except for a
depletion of \hi\ in the innermost regions), while in \eagle\ they are
significantly larger. Indeed, even considering a simulated galaxy
sample in which big \hi\ holes are not present, the
\citetalias{Gnedin2011} method systematically underpredicts the \hi\
surface density in the central region of the \eagle\ galaxies, while
equation (\ref{eq:Blitz}) -- with a slightly different choice of parameters
than in the present work \citep[see][]{Blitz2006} -- reproduces the
observations fairly well.

\begin{figure}
\centering
\includegraphics[width=0.43\textwidth]{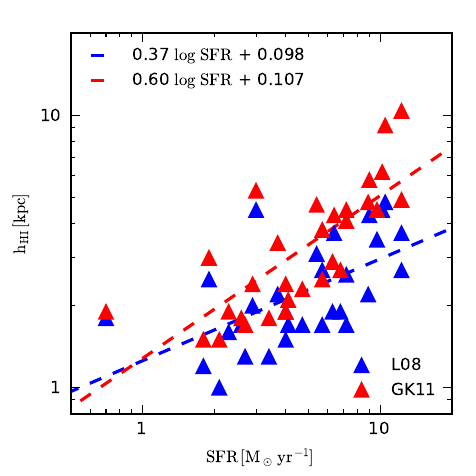}
\caption{\hi\ vertical thickness (defined as the height containing
  75\% of the \hi\ mass) as a function of the SFR within the galaxy
  $B$-band optical radii for the 30 Auriga haloes derived with the
  \citetalias{Leroy2008} (blue symbols) and \citetalias{Gnedin2011}
  (red symbols) prescriptions for the treatment of molecular
  hydrogen. Dashed lines represent the best-fitting relations in the
  two cases, with parameters indicated in the legend.}
\label{fig:hvssfr}
\end{figure}

\subsection{\hi\ vertical distribution}\label{sec:vertical}

We now focus on the vertical structure of our simulated \hi\ discs. As it
is apparent from the edge-on column density maps presented in
Section \ref{sec:morphologies}, the vertical extent of the \hi\ layer
varies for each object. Here we want to analyse this variation in a
more quantitative way and relate it to the internal properties of the
central galaxy, in particular to its SFR. There are,
of course, processes external to the galaxy that will also affect the
vertical \hi\ distribution, such as mergers (see e.g.~the cases of Au
11 and Au 30) or close interactions with satellites (e.g.~Au 18), which
however we do not discuss in the present work \citep[but see][for
an analysis of the impact of these effects on the vertical motion of gas
and stars within galactic discs]{Gomez2016b, Gomez2016a}.

It is worth recalling that in our galaxy formation physics model,
galactic outflows are powered by supernova explosions. The energy of
these outflows is related to the global SFR of the
galaxy. It is thus natural to expect a relation between this property
and the structural properties of the \hi\ disc, such as for instance
the vertical thickness of the \hi\ layer. Fig.~\ref{fig:hvssfr} presents the
corresponding correlation obtained with the two different methods we
used to compute the molecular hydrogen mass fraction (blue symbols for
the \citetalias{Leroy2008} and red symbols for the
\citetalias{Gnedin2011} prescription, respectively) together with
their best-fitting relations (dashed lines of the same colours). The
best-fitting parameters are given in the legend. To select the actively
star forming region of the galaxy, we consider only the region of the
disc within the optical radius of the galaxy $R_{\rm opt}$, computed
as the galactocentric radius at which the $B$-band surface brightness
profile reaches $25\, {\rm mag\,arcsec^{-2}}$. We then estimate the
total SFR of the galaxy and the \hi\ mass in a
cylinder of radius $R_{\rm opt}$, extending $\pm 30 \, \kpc$ outside
the disc plane. Finally, we define the \hi\ vertical thickness
$h_{\rm H{\sc I}}$ as the height containing 75\% of the \hi\ mass
determined in the step above.  The inferred values of
$h_{\rm H{\sc I}}$ for both the \citetalias{Leroy2008} and
\citetalias{Gnedin2011} methods are presented in
Table~\ref{tab:hiproperties}.

\begin{figure}
\centering
\includegraphics[width=0.45\textwidth]{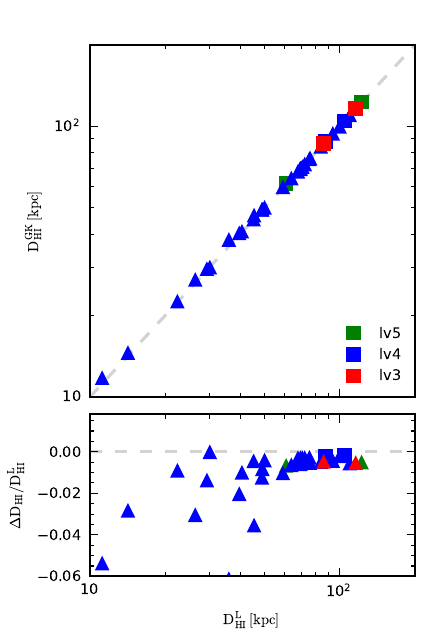}
\caption{\hi\ diameter mass obtained by using the
  \citetalias{Leroy2008} and the \citetalias{Gnedin2011} prescription
  to account for the molecular gas for the 30 Auriga haloes. Top
  panel: comparison of the \hi\ diameters obtained in the two
  cases. Bottom panel: fractional variation with respect to the
  \citetalias{Leroy2008} \hi\ diameter as a function of the
  \citetalias{Leroy2008} \hi\ diameter. Dashed lines in both panels
  show the 1:1 correspondence.}
\label{fig:diamcomp}
\end{figure}

\begin{figure}
\centering
\includegraphics[width=0.45\textwidth]{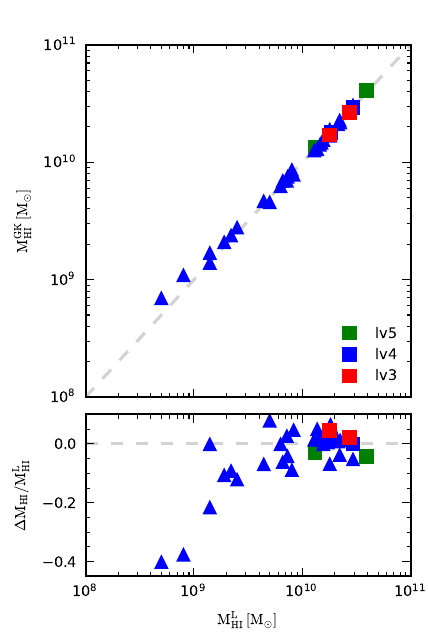}
\caption{As in Fig.~\ref{fig:diamcomp}, but for the inferred \hi\ mass
  of the Auriga sample.}
\label{fig:masscomp}
\end{figure}

Fig.~\ref{fig:hvssfr} shows a correlation between the \hi\ vertical
thickness and the SFR of the central galaxy, with more actively star
forming galaxies having larger values of $h_{\rm H{\sc I}}$. Overall,
this relationship between the star forming properties of the galaxy
and the thickness of its \hi\ layer is consistent with the
theoretical picture that galactic fountain flows generated by stellar
feedback can push a significant fraction of the \hi\ mass outside the
galaxy mid-plane \citep[see e.g.][]{Fraternali2006, Fraternali2008,
  Marasco2011}. This is also in agreement with the so-called
upside-down disc formation scenario in which a declining SFR leads to
a gas disc getting thinner with time \citep[see][for a detailed
analysis of this in a subsample of the Auriga haloes]{Grand2016a}. The degree of
scatter around the best-fitting relation is significant, supporting
the idea that also external processes contribute to building of the \hi\
layers. The correlation is present regardless of the
technique used to estimate the molecular gas fractions in the
simulations, but it is actually steeper -- and the associated \hi\
vertical thickness is larger -- for the \citetalias{Gnedin2011} method.
\FM{This can be explained the fact the \citetalias{Gnedin2011} model predicts in 
the central regions of the majority of the simulated galaxies a deficiency in 
the \hi\ content with respect to the \citetalias{Leroy2008} model. However, the 
total \hi\ masses are rather similar between the two methods (see 
Fig.~\ref{fig:masscomp}), which implies that in the \citetalias{Gnedin2011} case 
more \hi\ mass is found at larger vertical distances from the galaxy mid-plane.}

\FM{Unfortunately, from the observational point of view there are not many 
examples for which there exists a robust estimate of the vertical scale-height 
of the extended \hi\ layers. A list of edge-on galaxies for which this measure is 
available can be found in \citet[][and references therein]{Zschaechner2015}. For 
this set of objects, scale-heights are typically in the range $\sim 0.2-2.5$ 
kpc. Before comparing these values with the Auriga findings,  we caution the 
reader that there are different conventions for the \hi\ scale-height used in 
observations, which depend on the functional form fitted to the vertical \hi\ 
distribution and the use of one or multiple components in the fit, and those 
differ from the definition of vertical thickness that we have adopted in our work.}

\FM{Keeping in mind these caveats, we can examine how the trends inferred from 
Auriga compare to two prototypical examples of galaxies in which a HI halo has 
been detected: the Milky Way and NGC 891. In the Milky Way, \citet{Marasco2011} 
estimated a scale height of $1.6$ kpc for a SFR of $\sim3\,{\rm 
M_{\odot}\, yr^{-1}}$ \citep{Diehl2006}, which is compatible with what we find 
in our analysis. The same applies to NGC 891, which has a scale height of about 
$2$ kpc \citep{Oosterloo2007} for a total SFR of $3.8\,{\rm 
M_{\odot}\,yr^{-1}}$ \citep{Popescu2004}.}

\subsection{Differences in the global \hi\ properties}\label{sec:massdiff}

From the profiles computed in Section \ref{sec:profiles} it is possible to derive 
the total \hi\ mass of the galaxy as the mass of all the gas cells within 
$R_{\rm HI}$. We report these values in Table~\ref{tab:hiproperties} for all the 
30 objects we have simulated \footnote{Data are also available in electronic 
format at \url{http://auriga.h-its.org/data.html}}, together with their \hi\ radii and the 
average \hi\ surface density within $R_{\rm HI}$, for the two different methods 
we have used to remove the contribution of molecular hydrogen.

Figs \ref{fig:diamcomp} and \ref{fig:masscomp} illustrate the
differences between the recovered \hi\ diameters and \hi\ masses,
respectively, for both the \citetalias{Leroy2008} and
\citetalias{Gnedin2011} methods to account for molecular hydrogen. The
top panels of both figures present a direct comparison between the two
methods, while the bottom panels show their fractional variation with
respect to the \citetalias{Leroy2008} value. Different colours show
simulations at different resolution levels (see
Section \ref{sec:resolution} for more details), while the dashed lines
indicate the 1:1 correspondence.

Let us first focus on $D_{\rm HI} \equiv 2 R_{\rm HI}$. We have
already mentioned in Section \ref{sec:profiles} that at large radii
($R \sim R_{\rm HI}$) the method used to compute the
molecular fraction has a negligible impact on the results. Therefore,
it is not surprising that the \hi\ diameters estimated with the two
methods agree so well. Basically, all the simulated points lie very
close to the 1:1 expectation. The fractional difference between the
two $D_{\rm HI}$ values is of the order of a few per cent. The general
outcome is that the \citetalias{Gnedin2011} method tends to recover
slightly larger values for the \hi\ diameter. There is no obvious
trend of the fractional differences as a function of the \hi\ size of
the simulated objects, although it seems that they are more
pronounced at smaller $D_{\rm HI}$.

The same conclusions can be reached for the \hi\ masses, although here
the differences are larger (of the order of 10\% and reaching up to
40\% for small masses). However, in this case the fractional
differences between the two masses are spread approximately evenly
around the 1:1 value. This can be explained by the fact that the \hi\
diameters recovered by the \citetalias{Gnedin2011} method are
consistently larger than their \citetalias{Leroy2008} counterparts,
which compensates for the smaller \hi\ column densities found in the
inner regions of the disc for the \citetalias{Gnedin2011}
method. Again, there is no obvious trend in the fractional differences
as a function of the \hi\ mass, although the differences become more
distinct at low $M_{\rm HI}$.

\begin{figure}
\centering
\includegraphics[width=0.45\textwidth]{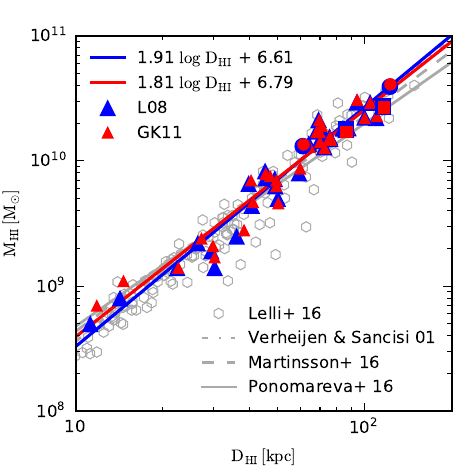}
\caption{The \hi\ mass diameter relation for the Auriga galaxies. Blue
  symbols show the results obtained with the empirical relation from
  \citetalias{Leroy2008} to account for molecular gas, while red
  symbols show the outcome of the prescription described in
  \citetalias{Gnedin2011} with lines of the same colours indicating the best-fitting
  relation inferred from the two samples. Gray lines and empty symbols, instead,
  compare the Auriga findings to observations of nearby galaxy as
  shown in the legend. Triangles present the results obtained at the
  standard (level 4) resolution, while circles and squares show haloes
  Au 16 and Au 24 at resolution levels 5 and 3, respectively \FM{a factor of 8 
  coarser or finer in mass resolution (see also Section {\ref{sec:resolution}})}.  }
\label{fig:massdiameter}
\end{figure}

\subsection{The mass--diameter relation}\label{sec:massdiam}

An important scaling relation in \hi\ observations of nearby disc
galaxies is the one linking the total \hi\ mass of the galaxy to the
size of the \hi\ disc, the so-called mass diameter relation
\citep{Broeils1997,Verheijen2001,Swaters2002, Noordermeer2005,
  Martinsson2016}. Fig.~\ref{fig:massdiameter} shows this relation for
our simulated sample of galaxies for the \citetalias{Leroy2008} (blue
symbols) and the \citetalias{Gnedin2011} (red symbols) methods to
account for the molecular hydrogen. \FM{Lines of the same colour} show
the best fit to this relation of the form
\begin{equation}
 \log\,\left(\frac{M_{\rm HI}}{{\rm M}_\odot}\right) = \alpha \log \, \left(\frac{D_{\rm HI}}{\kpc}\right) + \beta,
\end{equation}
with the best-fitting parameters given in the legend. \FM{Grey} lines 
indicate the best-fitting relations derived from
observations of nearby galaxies \citep{Verheijen2001, Martinsson2016,
  Ponomareva2016}. We also show, as empty symbols, observational results 
taken from the SPARC galaxy sample \citep{Lelli2016} to give an indication of 
the observed scatter in the relation. Our simulated galaxies reproduce the same 
general trend seen in observations, in which a larger \hi\ size corresponds to a 
higher \hi\ mass, and they agree with the observed scaling relation(s) (right 
panel) very well. \FM{The agreement is good even in terms of the rms scatter in 
the relation \citep[$\simeq 0.06$ dex in the residuals of the diameter distribution, see 
also][]{Wang2016}}. \FM{Reassuringly, the variations in the best-fitting 
relations inferred from the simulated \citetalias{Leroy2008} and 
\citetalias{Gnedin2011} samples are small, the \citetalias{Leroy2008} relation 
being slightly steeper. They are consistent with the observational uncertainties 
affecting the determination of the best-fitting parameters from real data.}

Our results on the mass-diameter relation are in agreement with the findings by 
\citetalias{Bahe2016}, who report that \eagle\ galaxies are consistent with the 
observational constraints. More precisely, \citetalias{Bahe2016} compare their 
simulated \hi\ discs to the \citet{Broeils1997} best-fitting relation which is 
not shown in Fig.~\ref{fig:massdiameter} but is equivalent (its slightly higher 
slope is compensated by a lower zero-point) to the ones considered in our 
analysis. Therefore, we conclude that, unlike the \hi\ surface density radial 
profiles, the mass-diameter scaling relation is a very robust outcome of 
simulations, even for very different treatments of the star-forming gas and 
stellar feedback. However, contrary to the \citetalias{Bahe2016} analysis, we do 
not observe a dramatic change in slope of the mass-diameter relation in the 
\citetalias{Gnedin2011} case. 

\FM{The above also implies that the average \hi\ surface density within $R_{\rm 
HI}$ is an approximately constant or slowly declining function of the \hi\ 
diameter in the Auriga galaxies in agreement with observations. Although there 
is considerably more scatter in this relation, its amplitude and the variability 
of the best fitting parameters are consistent with the observational findings.}

\begin{figure}
\centering
\includegraphics[width=0.45\textwidth]{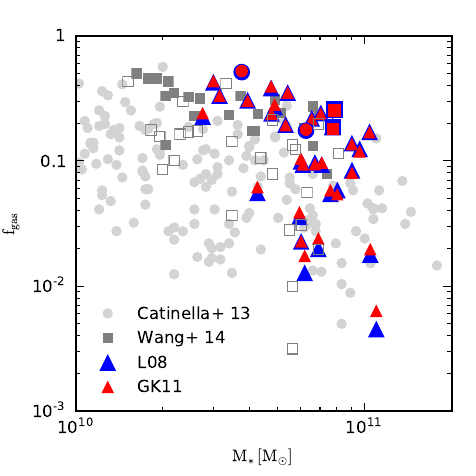}
\caption{\hi\ gas fraction as a function of the stellar mass for the
  Auriga sample. Stellar masses are computed within 10\% of the virial
  radius of each object, while \hi\ masses are those listed in Table
  \ref{tab:hiproperties}.  Results obtained through the empirical
  relation in \citetalias{Leroy2008} to account for molecular gas are
  shown with red symbols, while those obtained by applying the
  \citetalias{Gnedin2011} prescription are represented by blue
  symbols. Triangles give the results obtained at the standard
  (level~4) resolution, while circles and squares show haloes Au~16
  and Au~24 at resolution levels 5 and 3, respectively \FM{a factor of 8 
  coarser or finer in mass resolution (see also Section {\ref{sec:resolution}})}. Grey symbols
  indicate observational findings from the GASS \citep{Catinella2013}
  and the Bluedisk \citep{Wang2014} surveys, as indicated in the
  legend. For the combined data sets, only 195 out of 300 galaxies 
  which were significantly detected in \hi\ are shown.}
\label{fig:gasfrac}
\end{figure}

\subsection{Gas fractions}\label{sec:gasfrac}

In Fig.~\ref{fig:gasfrac} we present the \hi\ gas fractions
[$f_{\rm gas} \equiv M_{\rm HI} / (M_\star + M_{\rm HI})$] as a
function of the stellar mass for the Auriga sample. Stellar masses are
computed within $10\%$ of the virial radius of each object \citep[see
also][]{Marinacci2014a, Grand2016b}, and the \hi\ gas masses are those reported in
Table~\ref{tab:hiproperties}. Different colours indicate the two
different methods to compute the contribution of molecular hydrogen, as
indicated in the legend. The simulation results are compared \FM{to galaxies 
significantly detected in \hi} in both
the GALEX Arecibo SDSS Survey \citep[GASS,][]{Catinella2013} and Bluedisk \citep{Wang2013}
samples. The Bluedisk sample is further split between gas-rich
galaxies (full symbols) and a control sample (empty symbols), as in
\citet{Wang2013}.

Although the simulated range in stellar masses is somewhat reduced
compared to observations -- remember that the Auriga haloes were
selected to be a match for the Milky Way -- the Auriga set shows a
declining \hi\ gas fraction for increasing stellar mass as in the
observations. The decline in the gas fraction is comparable to the
observations, with an abrupt decrease at
$M_\star \sim 6\times 10^{10} {\rm M}_{\odot}$ where $f_{\rm gas}$
drops from $10\%$ to about $1\%$. These trends are robust with respect
to the model used to compute the molecular hydrogen fraction with
almost no difference in the recovered gas fractions, except at the
very high end of the stellar mass spectrum where, however, the
difference is within a factor of 2.

\begin{figure*}
\centering
\includegraphics[width=0.16\textwidth]{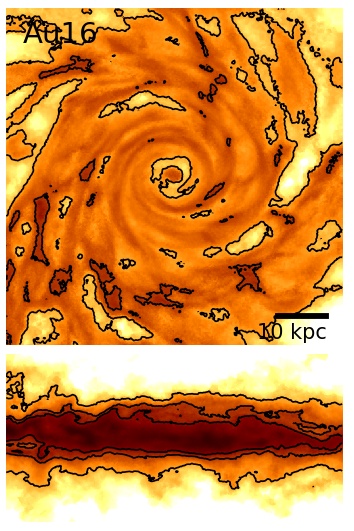}
\includegraphics[width=0.16\textwidth]{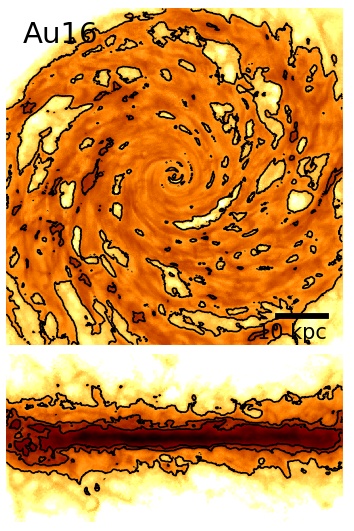}
\includegraphics[width=0.16\textwidth]{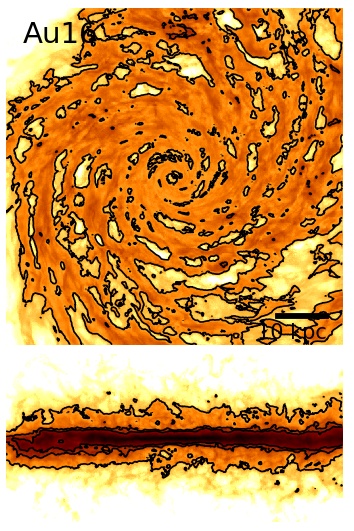}
\includegraphics[width=0.16\textwidth]{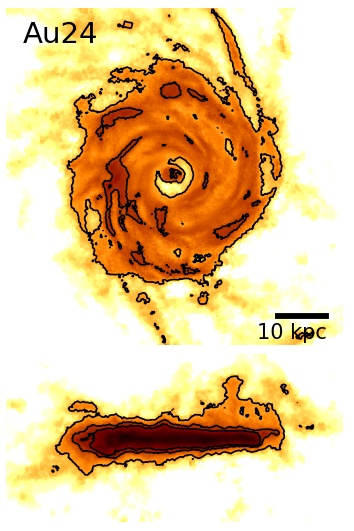}
\includegraphics[width=0.16\textwidth]{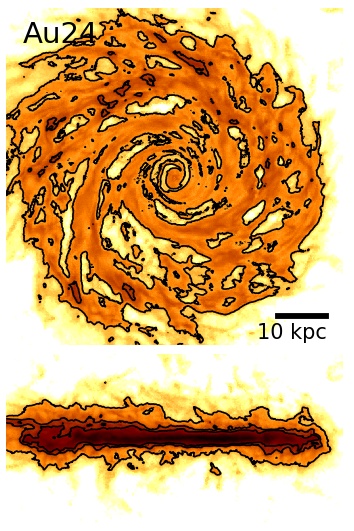}
\includegraphics[width=0.16\textwidth]{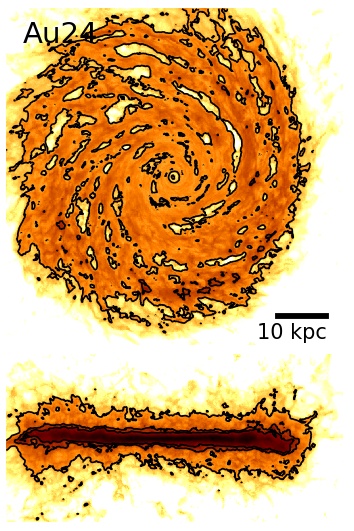}\\
\includegraphics[width=0.16\textwidth]{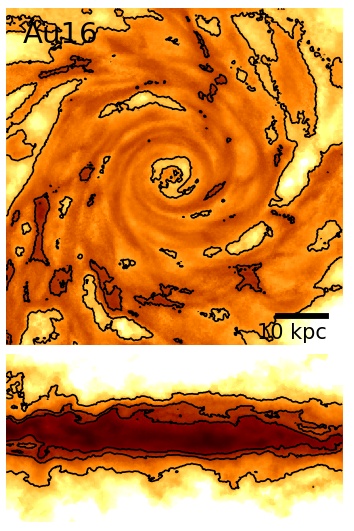}
\includegraphics[width=0.16\textwidth]{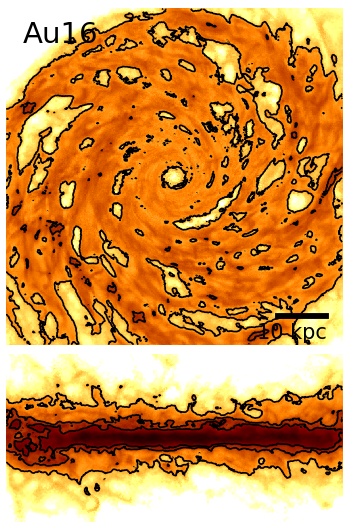}
\includegraphics[width=0.16\textwidth]{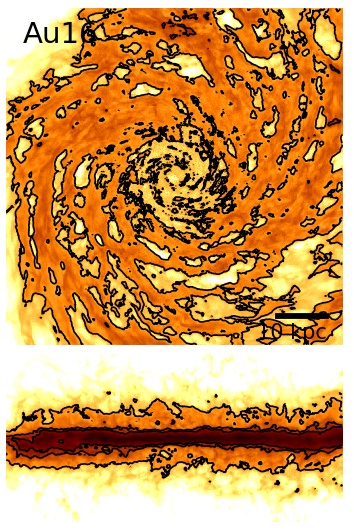}
\includegraphics[width=0.16\textwidth]{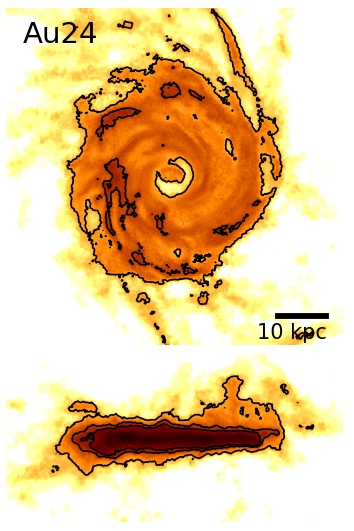}
\includegraphics[width=0.16\textwidth]{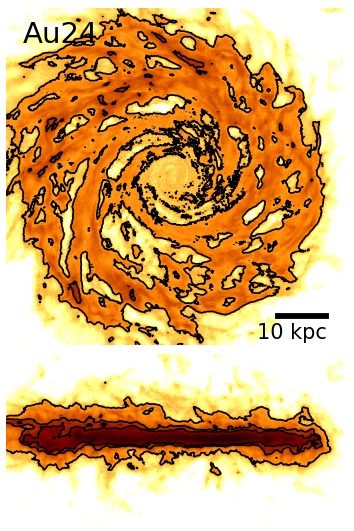}
\includegraphics[width=0.16\textwidth]{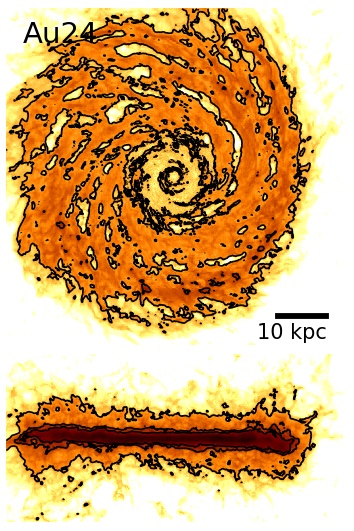}
\caption{Resolution study for the \hi\ column density maps of haloes
  Au~16 and Aq~24.  From left to right and for each halo, the
  resolution level varies from the least resolved to the most
  resolved.  The top row shows the maps obtained using the
  prescription from \citetalias{Leroy2008} for removing molecular gas,
  while the \hi\ maps in the bottom row adopt the \citetalias{Gnedin2011}
  approach.  The size of the projected region is 90 kpc on a side for
  face-on projections and $90\times 45$ kpc for the edge-on ones.}
\label{fig:HIprojres}
\end{figure*}

\FM{A comparison to the observations indicates that the Auriga galaxies} 
are systematically more gas-rich than the average for their stellar mass. If one 
considers the GASS sample, it can be readily seen that the Auriga set is at the 
upper envelope of the observed galaxies. If, instead, the Bluedisk sample is 
considered, the simulated galaxies follow more closely the trend of the gas-rich 
sample (which notably is composed of galaxies with \hi\ mass fractions 0.6 dex 
higher than the median relation found in \citealt{Catinella2010}) than the one 
of the control sample comprising more `average' galaxies. This occurs for 
stellar masses below the value $M_\star \sim 6\times 10^{10} {\rm M}_{\odot}$. 
As already discussed above, for stellar masses larger than this value, $f_{\rm 
gas}$ drops quite abruptly to the $1\%$ level. \FM{While being consistent with 
the gas fractions observed in galaxies detected in \hi, this value is not representative
of the global galaxy population given the numerous non-detections present in the two samples
that are not shown in the figure (about one third of the total targeted galaxies)}.

The fact that the Auriga galaxies are generally more gas-rich than the 
observations should not be surprising given the results already discussed in 
Sections~\ref{sec:profiles} and \ref{sec:massdiam}, in which we showed that our 
simulations overpredict the average surface density of the \hi\ discs and 
preferentially populate the high-mass, large-size part of the mass-diameter 
relation. Given that the stellar masses recovered by the Auriga project are 
consistent with abundance matching predictions \citep{Grand2016b}, \FM{we have 
investigated whether this higher \hi\ content might be due to the fact that the 
method presented in Section \ref{sec:neutrgas} does not apply any correction for 
photoionization (originating from the cosmic UV background and the local stellar 
radiation field) to the star-forming gas, thus overestimating its neutral gas 
fraction. The contribution of this gas phase to the total \hi\ content of the 
Auriga galaxies is highly variable, ranging from $\sim 10\%$ to 
being the predominant contribution to the HI mass ($\gsim 95\%$) in the most 
compact galaxies such as Au 26 and Au 13. On average the contribution of star-forming gas
to the total \hi\ mass is about 45\%. Therefore, applying a correction for photoionisation to the 
star-forming gas might alleviate the tension with the observations on the gas 
content of the Auriga galaxies, especially in the case of less extended \hi\ discs.} \FM{To further study 
this issue, it would also be informative to check whether the observational trends, 
including \hi\ non-detections, could be better reproduced by a more general 
population of simulated haloes not chosen to be a match for the Milky Way, in 
line with what has been achieved in uniformly-sampled cosmological boxes such as 
\eagle\ (\citetalias{Bahe2016}; \citealt{Crain2016}).}

\subsection{Resolution dependence}\label{sec:resolution}

Up to now we have used our fiducial resolution level to analyse the
\hi\ properties of the Auriga galaxies. A few selected galaxies in the
Auriga set have been re-simulated also at different mass and spatial
resolution levels. These additional simulations have been performed
with the same galaxy formation physics model as the fiducial runs;
only the number of resolution elements in the high-resolution region
(and consequently the force resolution of the simulations) has been
modified. In this section, we are interested in checking how robust
the results of our \hi\ analysis are with respect to a change in the
resolution of the runs.

We first introduce a naming convention for easy reference to runs
differing only in numerical resolution. Following the nomenclature
adopted in the Aquarius project \citep{Springel2008}, we identify each
resolution level with a numeral. A difference of one resolution level
means a finer (or coarser) mass resolution by a factor of 8, thus
implying a spatial resolution (i.e. a softening length) which can be
finer (or coarser) by a factor of 2. The lower the numeral identifying
the resolution level, the finer the effective (mass and spatial)
resolution. In what follows, we are going to analyse three different
resolution levels for the haloes Au~16 and Au~24: our fiducial runs at
resolution level 4, and finer (level 3) and coarser (level 5)
realizations.

Fig.~\ref{fig:HIprojres} presents \hi\
face-on and edge-on density projections (in a way akin to
Figs~\ref{fig:HIimages1}--\ref{fig:HIimages3}) for the haloes
Au~16 and Au~24. The top row in each figure shows the
\hi\ surface density obtained with the \citetalias{Leroy2008}
prescription to compute the contribution of molecular hydrogen, while
in the bottom row the \citetalias{Gnedin2011} model has been used. The
resolution increases for each halo from left to right such that each 
column in the figure represents \hi\ density projections of either halo 
Au~16 or Au~24 at the same resolution level but obtained with 
different treatments for molecular hydrogen.

From the figure, it can be seen that the main morphological traits of
the \hi\ discs are present for both haloes at all three different
resolution levels. Not surprisingly, higher resolution brings out
finer details in the gas distribution. In particular, in the edge-on
projections it is apparent that the thin component of the \hi\ disc is
better defined at the highest resolution level, while the vertical gas
distribution puffs up considerably at the lowest resolution. The central depletion 
in the \hi\ content for the \citetalias{Gnedin2011} method is
barely noticeable at the lowest resolution level for both haloes, but
it is increasingly more visible at higher resolution -- note how the
size of the central \hi\ `hole' increases in size for the Au~16
halo. The most striking difference, however, concerns the radial
extent of the \hi\ disc in the halo Au~24. Indeed, at resolution
level~5 the \hi\ disc is significantly less extended than in the two
higher resolution levels, where the size of the \hi\ disc appears
similar. Halo Au~16 is not affected by this discrepancy, and the
variation of the disc size across the different resolution levels is
smaller.

\begin{figure}
\centering
\includegraphics[width=0.42\textwidth]{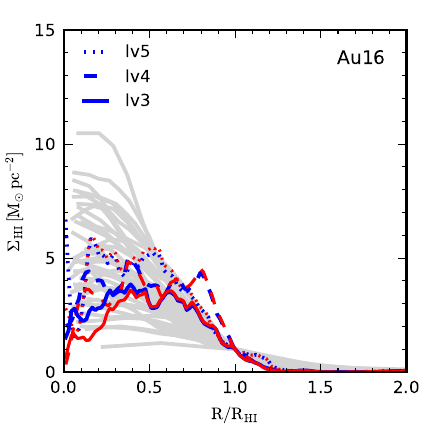}
\includegraphics[width=0.42\textwidth]{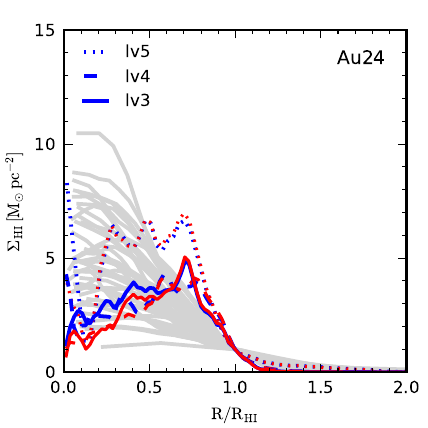}
\caption{Resolution study for the \hi\ surface density profiles for
  haloes Au~16 and Au~24. The blue and red lines indicate the profiles
  obtained using the prescription from \citetalias{Leroy2008} or
  \citetalias{Gnedin2011} for removing molecular gas,
  respectively. The line style indicates the resolution level, as
  given in the legend. Light grey lines display the \hi\ radial
  profiles of the Bluedisk survey \citep{Wang2014}.}
\label{fig:HIprofileres}
\end{figure}

Fig.~\ref{fig:HIprofileres} presents \hi\ surface density profiles
as a function of the galactocentric radius (normalized to
$R_{\rm HI}$) for haloes Au~16 (top) and Au~24 (bottom). Line colours
show the results obtained with the \citetalias{Leroy2008} (blue) and
\citetalias{Gnedin2011} (red) prescriptions to compute the molecular
hydrogen mass fraction, while their line-style identifies the
resolution level as indicated in the legend. Light grey lines are the
\hi\ surface density profiles of the Bluedisk survey and are shown for
comparison.

The \hi\ radial profiles show the general trends already discussed
in Section \ref{sec:profiles}, that is an increasing \hi\ surface density
profile in the innermost regions which reaches a maximum value or a
plateau and then quickly drops at $R \simeq 0.85\times R_{\rm HI}$,
leaving little gas past the \hi\ radius of the galaxy. 
The \hi\ surface density profiles
are overall very similar across the resolution levels. Only level 5
presents marked differences with respect the two finer resolution
levels, in particular for halo Au~24. In general, the profiles for
level~4 and level~3 track one another closely, regardless of the
prescription adopted to estimate the molecular gas fraction, although
for halo Au~16 the level 4 profile predicts an \hi\ column density
about a factor of 2 larger than the corresponding level~3 profile at
$R \sim 0.8 \times R_{\rm HI}$.  A comparison of the simulated
profiles with those obtained in the Bluedisk survey shows that the
profiles of the Auriga galaxies are consistent with the observed ones,
although occasionally (e.g. for halo Au~16 at level~4 and halo
Au~24 at level~5) there is an excess of \hi\ at
$R \simeq 0.8 \times R_{\rm HI}$.

Starting from the radial \hi\ density profiles it is possible to
derive the structural properties of the \hi\ discs (sizes, total
masses and average densities) at the additional resolution levels in
the same way as done for the fiducial resolution. We report these
values in Table~\ref{tab:hiproperties} under separate headings for the
different resolution levels. Clearly, it is possible to investigate
several scaling relation between these quantities, as already done in
Sections~\ref{sec:massdiff} to \ref{sec:gasfrac}, allowing us to check
the robustness of our findings with respect to resolution changes. To
this end, we added the relevant data points for haloes Au~16 and
Au~24 at resolution levels 5 and 3 to the plots presented
above, either in different colours (Figs.~\ref{fig:diamcomp} and
\ref{fig:masscomp}) or with different symbols (circles for level~5 and
squares for level~3; see Figs \ref{fig:massdiameter} and
\ref{fig:gasfrac}). Although differences between the sizes, masses and
average densities exist across the three resolution levels, the
individual data points follow the general pattern determined by the
full set of 30 simulations at resolution level~4, meaning that each
individual galaxy follows the general scaling relations derived for
the full sample.

From these considerations we can conclude that our findings are robust
with respect to changes in mass resolution (over a factor of 64) and this
holds regardless of the prescription adopted to estimate the
contribution of the molecular gas to the total neutral hydrogen
density. While some discrepancies do exist between the different
resolution levels, they tend to decrease with increasing
resolution. In particular, our fiducial resolution level seems to be
adequate to model the \hi\ content of Milky Way-like systems with
sufficient accuracy.

\section{Summary and conclusions}\label{sec:conclusions}

In this paper we have analysed the properties of the \hi\ distribution
in the Auriga simulations, a set of cosmological MHD
simulations run with the moving-mesh code \arepo\ that investigate the
formation and evolution of disc galaxies similar to our own Milky
Way. We have modelled the \hi\ gas by computing the neutral hydrogen
fraction separately for low-density and star-forming gas, which, in
the star formation module currently implemented in our simulations, is
described by an effective equation of state and thus requires a
special treatment to calculate its neutral fraction in a meaningful
way. To estimate the contribution of molecular gas to the total
neutral hydrogen mass we have used two different prescriptions --
motivated by observational and theoretical studies -- and compared
their outcomes. Our main results can be summarized as follows.

\begin{enumerate}
\item The vast majority of the simulated systems feature an extended
  (i.e. more extended than the associated stellar distribution) \hi\
  disc which in some cases may reach a radial extent of
  $\sim 60\,\kpc$. A few objects (i.e. Au~11 and Au~30) present
  disturbed morphologies as a consequence of late-time galaxy
  interactions.
 
\item The vertical extent of the \hi\ distribution varies on a system
  by system basis.  The amount of \hi\ gas outside the disc plane
  correlates with the amount of star formation occurring in
  the (stellar) disc, in agreement with a scenario where most of the
  extra-planar \hi\ gas results from a fountain-like gas flow
  \citep{Fraternali2006,Fraternali2008,Marasco2011}.
 
\item The (face-on) radial profiles of the \hi\ surface density show a common 
  pattern in which the \hi\ surface density stays approximately constant and then 
  exponentially drops at $\sim R_{\rm HI}$. Contrary to previous theoretical 
  studies \citepalias{Bahe2016}, \FM{which however employed a substantially lower resolution than 
  Auriga}, our simulated galaxies do not show a deficiency 
  of \hi\ in the central region, but rather overpredict the \hi\ contents at 
  galactocentric radii $R\simeq 0.5\,R_{\rm HI}$.

\item Our simulated \hi\ discs lie on the observed mass-diameter
  relation \citep[e.g.][]{Broeils1997}. The scatter in the
  mass-diameter relation and the distribution of the average \hi\
  surface density within the \hi\ radius are also consistent with
  those derived for a number of observational data sets
  \citep{Verheijen2001, Lelli2016, Martinsson2016, Ponomareva2016}.

\item The \hi\ gas content in the Auriga galaxies, expressed in terms
  of gas fractions, is a declining function of the stellar mass
  present in the system, in agreement with observational
  findings. Compared to observations, our \hi\ fractions are at
  the upper end of the allowed observational range, indicating
  that our simulated systems are systematically more gas-rich than
  typical nearby galaxies.

\item The main results of our analysis are robust with respect to
  changes in the prescription used to estimate the contribution of
  molecular hydrogen to the total neutral hydrogen mass, although the
  use of the \citetalias{Gnedin2011} model yields \hi\ surface densities
  that are lower in the central regions than those estimated via the
  empirical \hi/H$_2$ mass ratio relation of \citetalias{Leroy2008}.
 
\item Our findings are slightly sensitive to variations in numerical
  resolution, but we note that the results for the two finer
  resolution levels are in good agreement with one
  another. Notwithstanding that, even the simulations run at the
  coarsest resolution level result in systems having \hi\
  distributions whose properties are consistent with the observational
  trends.
\end{enumerate}

Our analysis exemplifies the significant progress achieved by
cosmological simulations of galaxy formation -- and especially with
respect to the formation of galaxies similar to our own Milky Way --
which can now not only reproduce important structural properties (such
as stellar masses, disc scale lengths and SFRs) and
scaling relations associated with the stellar component of late type
systems, but are also able to make meaningful predictions for the
(neutral) gas component within the main galaxy. The predictions of
these properties can be used, in conjunction with an accurate
comparison to observations, to put new and more stringent constraints
to the sub-grid physics used in these calculations, which is one of
the major sources of uncertainties in current cosmological
simulations.

In addition, having simulated gaseous discs whose properties are in
broad agreement with the observed trends in nearby disc galaxies, one
can now pursue interesting new lines of research. One promising
direction in the context of the \hi\ gas would be to carry out
detailed studies of its dynamical state. This is an already well
established observational field in radio astronomy for which many
techniques have been developed and tested. The same techniques can
now be systematically applied to simulation data, likely yielding
informative comparisons between the simulated galaxies and the
observations, an avenue that we intend to explore in future work.

\section*{Acknowledgements}

We thank an anonymous referee for a constructive report. 
FM also thanks A. Ponomareva for helpful discussions and for kindly
providing the best-fitting parameters of the observed \hi\
mass--diameter relationship, and F. Fraternali for useful comments. 
FM is grateful for the hospitality of
the Heidelberg Institute for Theoretical studies during the final
stages of this work. RG and VS acknowledge support by the DFG Research
Centre SFB-881 `The Milky Way System' through project A1. This work
has also been supported by the European Research Council under ERC-StG
grant EXAGAL- 308037. Part of the simulations of this paper used the
SuperMUC system at the Leibniz Computing Centre, Garching, under the
project PR85JE of the Gauss Centre for Supercomputing. All the figures
in this work were produced by using the {\sc matplotlib} graphics
environment \citep{Matplotlib}.

\bibliographystyle{mnras}
\bibliography{paper}

\label{lastpage}

\end{document}